\documentclass[11pt]{article}

\usepackage[final]{acl}

\usepackage[utf8]{inputenc}
\usepackage[T1]{fontenc}
\usepackage{times}
\usepackage{latexsym}
\usepackage{microtype}
\usepackage{inconsolata}

\usepackage{amsmath}
\usepackage{amssymb}
\usepackage{amsfonts}
\usepackage{bm}
\usepackage{mathtools}

\usepackage[ruled,vlined]{algorithm2e}

\usepackage{graphicx}
\usepackage{hyperref}
\usepackage{booktabs}
\usepackage{multirow}
\usepackage{makecell}
\usepackage[table]{xcolor}

\usepackage{natbib}

\usepackage{enumitem}
\usepackage{verbatim}
\usepackage{amsthm}
\usepackage{mathrsfs}
\usepackage{subfigure}

\definecolor{brightblue}{rgb}{0.0, 0.0, 1.0}

\title{TIEM: Temporal Integration of Hypergraph Evidence and Skill Memory for Event-Driven Financial Forecasting}

\author{
\textbf{Wenjin Liu}\textsuperscript{1,2,*} \quad
\textbf{Shen Pang}\textsuperscript{2,*} \quad
\textbf{Chenxi Wang}\textsuperscript{2} \quad
\textbf{Tiesunlong Shen}\textsuperscript{3} \quad
\textbf{Jiajie He}\textsuperscript{4} \\
\textbf{Zhe Cui}\textsuperscript{2,$\dagger$} \quad
\textbf{Xiaobao Wu}\textsuperscript{5} \quad
\textbf{Anh Tuan Luu}\textsuperscript{1} \quad
\textbf{Haoran Luo}\textsuperscript{1,$\dagger$} \\
\textsuperscript{1}Nanyang Technological University \quad
\textsuperscript{2}Hithink Research \quad
\textsuperscript{3}National University of Singapore \\
\textsuperscript{4}University of Maryland, Baltimore County \quad
\textsuperscript{5}Shanghai Jiao Tong University \\
\texttt{wenjinliu23@outlook.com, haoran.luo@ntu.edu.sg} \\
\href{https://qwenqking.github.io/Fin_TIEM/}{%
\raisebox{-0.2ex}{\includegraphics[height=1em]{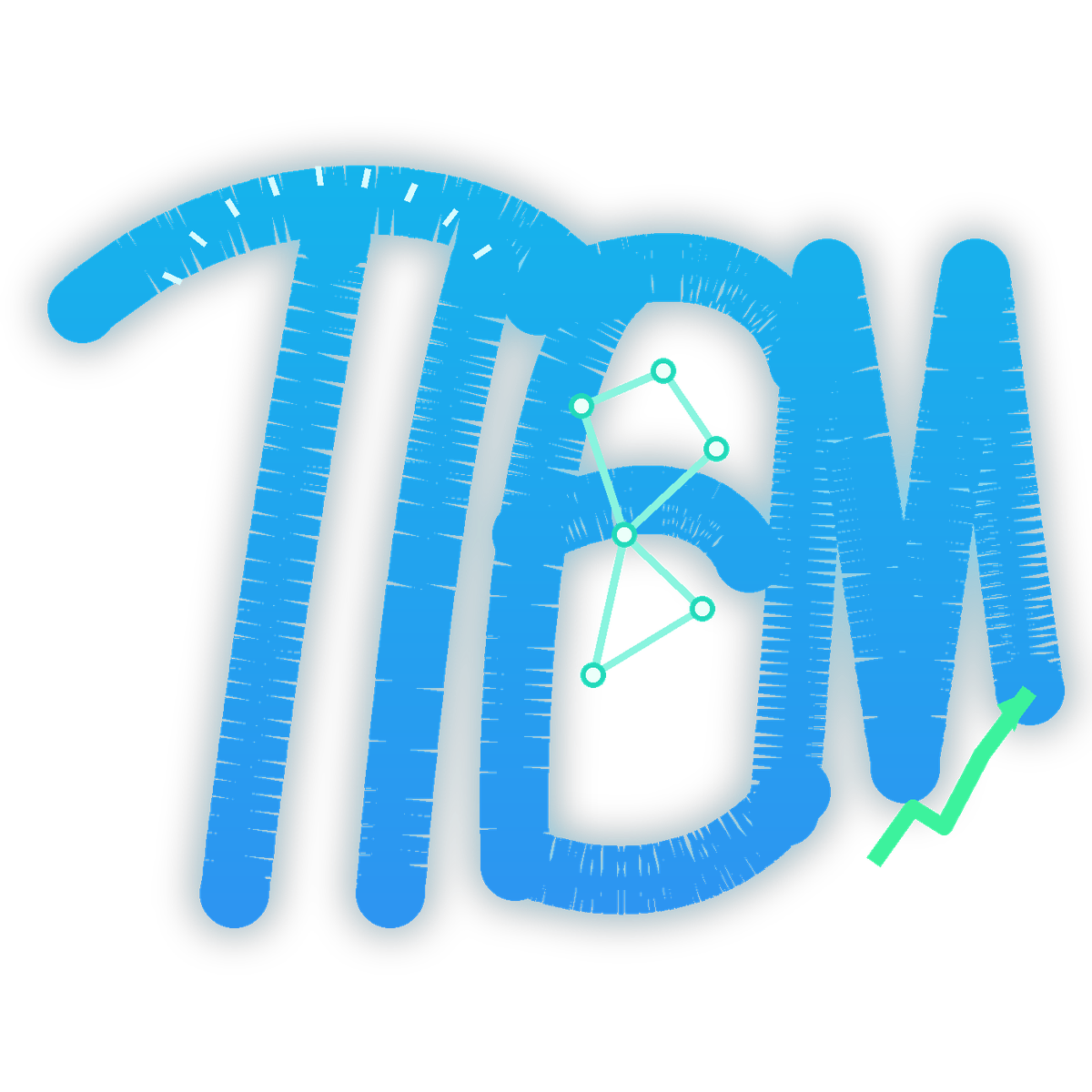}}\;Homepage%
} \quad
\href{https://github.com/QwenQKing/Fin_TIEM}{%
\raisebox{-0.2ex}{\includegraphics[height=1em]{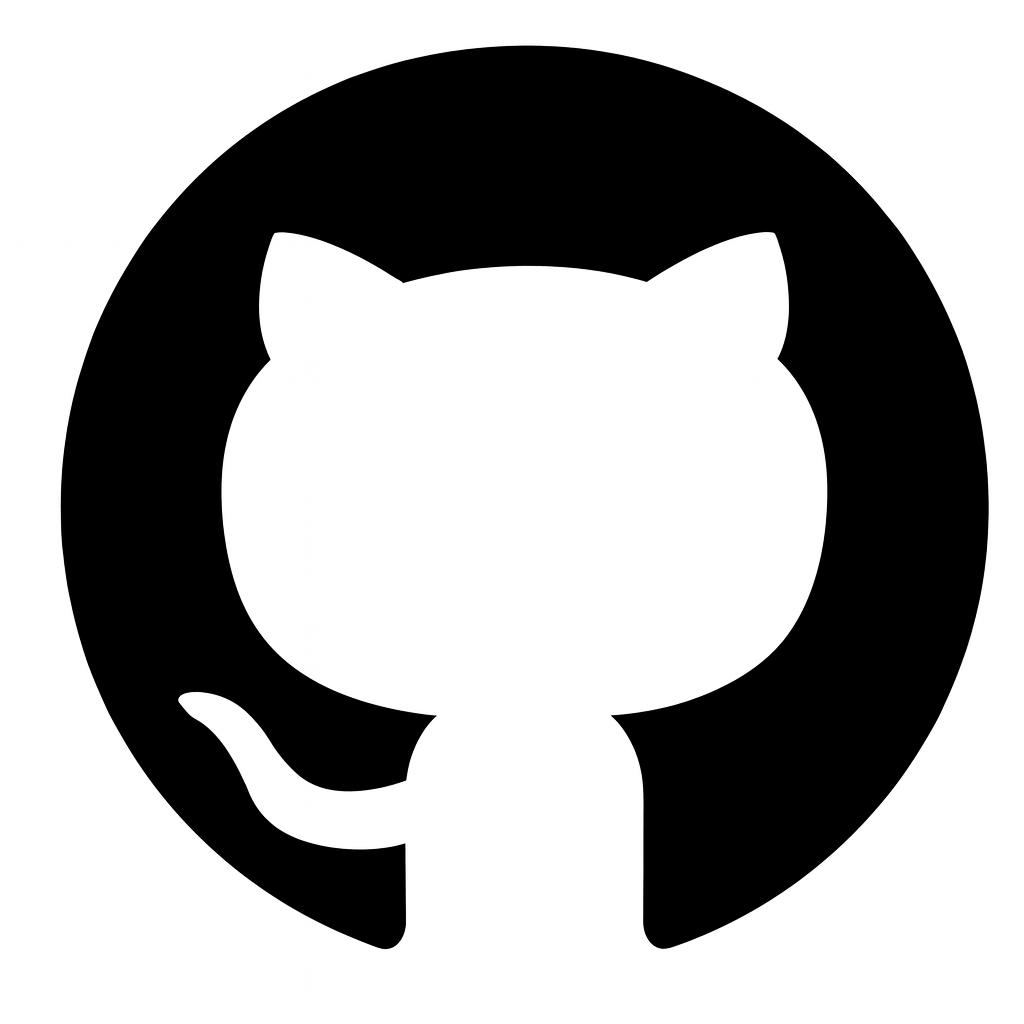}}\;GitHub%
} \quad
\href{https://huggingface.co/datasets/QwenQKing/TIEM-dataset}{%
\raisebox{-0.2ex}{\includegraphics[height=1em]{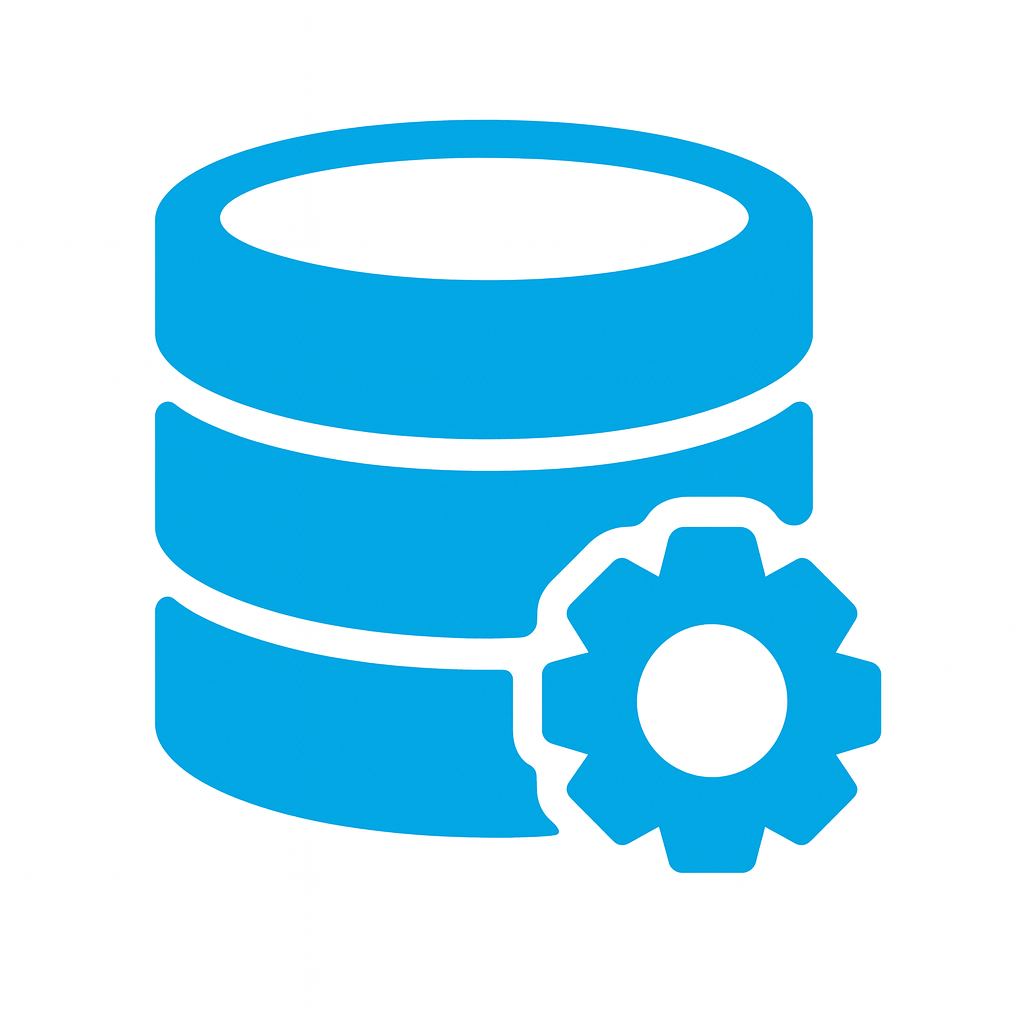}}\;Dataset%
} \quad
\href{https://huggingface.co/datasets/QwenQKing/TIEM-databases}{%
\raisebox{-0.2ex}{\includegraphics[height=1em]{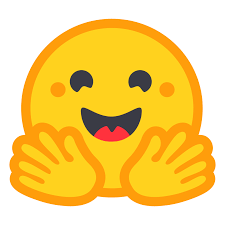}}\;Huggingface%
}
}

\begin{document}
\maketitle
\begingroup
\renewcommand{\thefootnote}{}
\footnotetext{\textsuperscript{*} Equal contribution. \textsuperscript{$\dagger$} Corresponding authors.}
\endgroup

\begin{abstract}
Event-driven catalyst-outcome forecasting increasingly uses retrieval- and
memory-augmented large language model agents for prediction. However, training-data
contamination and temporal leakage can create an Evidence Chasm
between reported accuracy and true predictive ability. We propose
\textbf{TIEM}, a timestamp-gated framework with three
coordinated components: an Event-Evidence Hypergraph (EEH) for
timestamp-filtered multi-tier retrieval; a Case-based Skill Memory
(CSM) for source-tagged temporal skills; and
Heterogeneous Evidence-Experience Fusion Reasoning (HEFR) for
evidence-experience fusion and prediction.
We also introduce FinPURE, a recent-period A-share holdout
benchmark and use a Name-Date Probe to assess per-model
name--date sensitivity rather than assuming training cutoffs.
Results on five financial forecasting benchmarks show \textbf{TIEM} outperforms current baselines. Our project is available\footnote{\url{https://github.com/QwenQKing/Fin_TIEM}}.
\end{abstract}

\section{Introduction}

Event-driven catalyst-outcome forecasting infers price moves
from catalyst-event text, a financial application of large language models
(LLMs)~\cite{lee2025hybgrag,jin2025disentangling}. Research has evolved from price-based time-series
models~\cite{uddin2025unseentimeqa,tan2026artem} and announcement sentiment
analysis~\cite{li2025investorbench} to retrieval-augmented generation
(RAG)~\cite{man2025context} and LLM agents with evolvable
memory~\cite{mem0, a-mem}, improving event understanding, outcome
prediction, and decision support. However, training-data
contamination~\cite{tang2025anre,tan2025prospect} and temporal leakage~\cite{ouyang2025hoh,xu2025triplefact}
can widen the Evidence Chasm between reported scores and genuine
forecasting ability, making benchmark scores unreliable proxies for
real-world predictive performance in practical financial applications~\mbox{\cite{qiu2025text,zhao2025uncertainty}}.

\begin{figure}[t]
    \centering
    \includegraphics[width=1\linewidth]{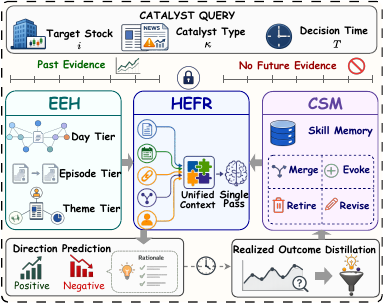}
    \caption{An illustration of the \textbf{TIEM} framework.}
    \label{fig:framework}
\end{figure}

\begin{figure*}[t]
\centering
\includegraphics[width=0.98\textwidth]{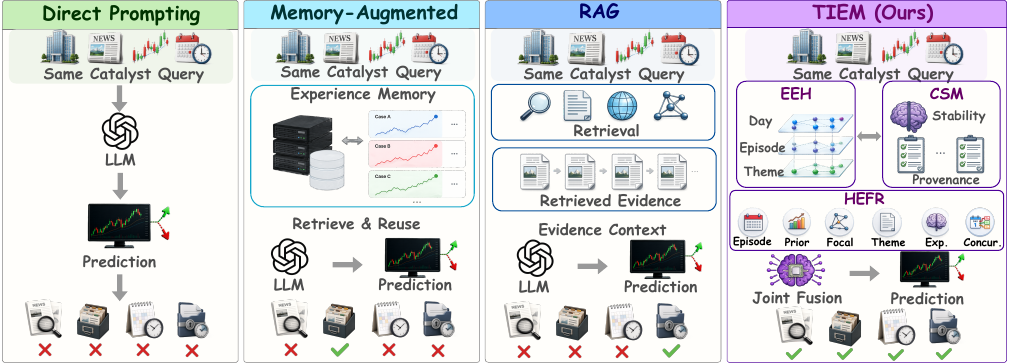}
\caption{Comparison of direct prompting, memory augmentation, RAG, and \textbf{TIEM} on shared test instances.}
\label{fig:baselines-related}
\end{figure*}

For event-driven financial forecasting, a series of methods
have been proposed~\cite{Li2026JanusQ,rajesh2026beyond,huang2026mem,tran2026prime,lin2025fact,du2026memguide}. Retrieval-augmented methods
(GraphRAG~\cite{graphrag}, HyperGraphRAG~\cite{hypergraphrag},
HippoRAG~\cite{hipporag}, LightRAG~\cite{lightrag}) build graph- or
hypergraph-structured indexes over financial corpora and retrieve,
via similarity queries, historical evidence relevant to the
catalyst event for fact-grounded prediction; memory-augmented
systems (Mem0~\cite{mem0}, A-MEM~\cite{a-mem}, MemGPT~\cite{memgpt})
accumulate past interactions in persistent memory and inject
historical experience into the LLM's reasoning context through
retrieve-and-reuse~\cite{tang2025evowiki}.

However, these methods still face three key challenges.
\mbox{\textit{(i)} \textbf{Multi-scale temporal structure.}} Catalyst events span
same-stock event chains and cross-industry thematic co-movements, but
flat single-timestamp retrieval loses multi-scale structure and admits
future information, amplifying leakage risk.
\mbox{\textit{(ii)} \textbf{Experience stability and source metadata.}} Transferable
knowledge needs stable records with source metadata, but flat memory
lacks such metadata or timestamp-gated write constraints, aggravating
contamination.
\mbox{\textit{(iii)} \textbf{Heterogeneous evidence fusion.}} Catalyst outcomes depend
on coupled focal events, same-stock priors, episodes, cross-stock
themes, and historical experience, but existing methods flatten or stage
them, breaking the multi-source joint context. Together, they weaken
forecast quality.

To address these challenges, we propose \textbf{TIEM}
(as shown in Fig.~\ref{fig:framework}), a \textbf{timestamp-gated} framework
that realizes temporal integration through three coordinated components
over a time-stratified hypergraph and case-based skill memory, alongside
\textbf{FinPURE}, a recent-period A-share holdout benchmark with
per-model name--date sensitivity assessed by the probe. First, EEH
organizes atomic event facts and higher-level records with
explicit temporal extents for timestamp-filtered multi-tier retrieval.
Furthermore, CSM maintains source-tagged skills via EVOKE, REVISE,
MERGE, and RETIRE primitives and supplies a retrieved skill stream.
In addition, HEFR fuses five budgeted streams (focal, prior, episode,
theme, experience) plus same-day concurrent facts in one prompt.

To evaluate TIEM, we conduct a series of comparative experiments on five diverse
event-driven catalyst-outcome benchmarks. Experimental results show that TIEM
outperforms the current baselines (as illustrated in Fig.~\ref{fig:baselines-related})
across the evaluated in-distribution (ID), cross-dataset, and recent-period
settings, supporting timestamp-gated evidence integration for event-driven
catalyst-outcome forecasting under explicit temporal constraints.

\begin{figure*}[t]
\centering
\includegraphics[width=0.98\textwidth]{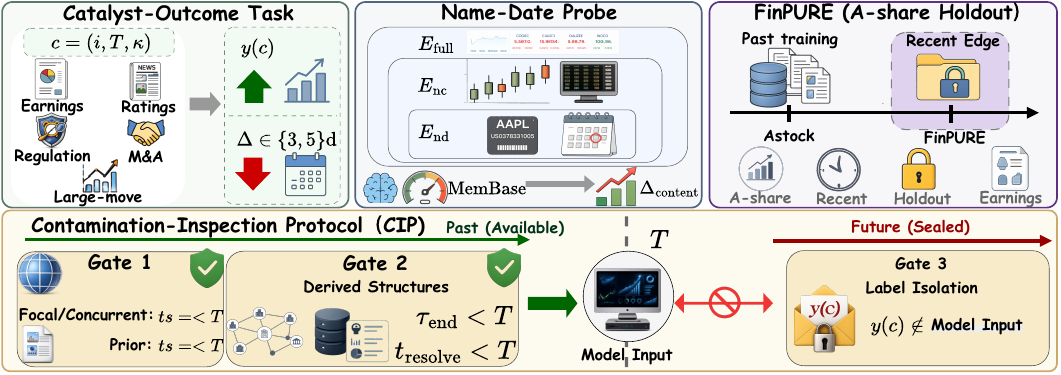}
\caption{Overview of the catalyst-outcome task, Name--Date Probe, FinPURE A-share holdout, and three-gate CIP.}
\label{fig:baselines-1}
\end{figure*}

\section{Related Work}\label{sec:related}

\textbf{Retrieval-Augmented Generation.}
Beyond vanilla RAG~\cite{rag-lewis}, work extends retrieval in three
directions~\cite{guo2025dior,jin2025hierarchical,dong2025rag,xin2025sparse,chen2025drag,yan2025rpo}. For retrieval control, RAPTOR~\cite{raptor} builds a
hierarchical tree index through recursive abstractive summarization;
Self-RAG~\cite{selfrag}, CRAG~\cite{crag}, and
Adaptive-RAG~\cite{adaptive-rag} introduce self-evaluation and
adaptive triggering. For structured indexing, GraphRAG~\cite{graphrag},
LightRAG~\cite{lightrag}, and HyperGraphRAG~\cite{hypergraphrag} model
entities and relations as graphs or hypergraphs, supporting multi-hop
reasoning. T-GRAG~\cite{t-grag},
GenTKG~\cite{gentkg}, and
STAR-RAG~\cite{star-rag} address time-evolving knowledge;
A-RAG~\cite{arag2026} couples retrieval with reasoning.

\textbf{Memory-Augmented and Skill-Evolving LLM Agents.}
Persistent memory enables LLM agents to accumulate
experiential knowledge~\cite{luo2026storage, du2026memorysurvey}.
Operation-centric systems (Mem0~\cite{mem0}, A-MEM~\cite{a-mem},
Zep and Graphiti~\cite{zep-graphiti}) manage memory through
standardized add, retrieve, and update primitives. Graph-structured
memory like HippoRAG~\cite{hipporag, hipporag2} simulates hippocampal
episodic recall via PageRank. For domain customization,
FinMem~\cite{finmem} designs layered cognitive memory for financial
decisions; MADAM-RAG~\cite{madam-rag-conflicting-evidence} handles
conflicting evidence. For continual skill evolution,
AutoSkill~\cite{yang2026autoskill}, MemSkill~\cite{zhang2026memskill},
MUSE-Autoskill~\cite{lin2026muse}, and SkillRL~\cite{xia2026skillrl}
distill cumulative experience into reusable skills, systematically
benchmarked by Evo-Memory~\cite{evomemory2025} and SSGM~\cite{ssgm2026}.

\begin{figure*}[t]
\centering
\includegraphics[width=1\textwidth]{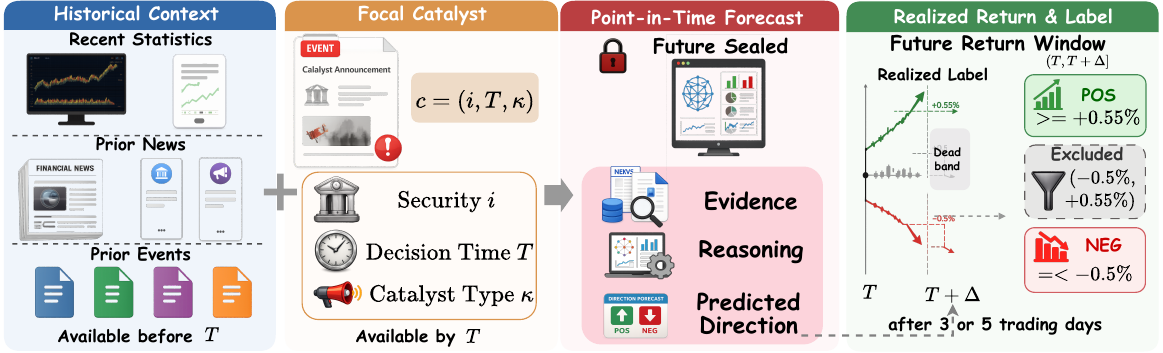}
\caption{Timestamp-gated catalyst-outcome forecasting from evidence to direction prediction and realized labels.}
\label{fig:baselines}
\end{figure*}

\section{Task, Benchmark, and Protocol}
\label{sec:task}

\noindent\textbf{Task Definition.}
Catalyst-Targeted Reframing (CTR) defines each forecast as
$c=(i,T,\kappa)$, where $i$, $T$, and $\kappa$ are the security, decision
time, and catalyst type. Labels use
$y(c)=\mathrm{sign}\,r_{i,\Delta}(T)$ after discarding returns in the
$(-0.5\%,+0.55\%)$ deadband, where $r_{i,\Delta}(T)$ is the
$\Delta$-trading-day return and $\Delta\!\in\!\{3,5\}$ is
benchmark-specific. NDP compares nested inputs
$E_{\mathrm{nd}}\subset E_{\mathrm{nc}}\subset E_{\mathrm{full}}$
containing, respectively, security/date, those fields plus recent
statistics and prior headlines, and those inputs plus focal-event text.
On held-out data, predictor $f$ yields
$\mathrm{MemBase}=\mathrm{Acc}(f;E_{\mathrm{nd}})$ and
$\Delta_{\mathrm{content}}=\mathrm{Acc}(f;E_{\mathrm{full}})-
\mathrm{Acc}(f;E_{\mathrm{nc}})$.

\noindent\textbf{Benchmark Construction.}
FinPURE is an earnings-only, recent-period A-share holdout for temporal
generalization without an unseen catalyst type. Astock supplies
in-distribution evaluation, CMIN-US and EDT cross-market evaluation, and
CSMD supplies cross-time evaluation for temporal shift, testing transfer to a later period.

\noindent\textbf{Contamination Inspection Protocol.}
CIP screens temporal and shortcut risks without a universal training cutoff. Decision
time $T$ follows each dataset's timestamp granularity, with date-only records available
at end of day. Focal and concurrent evidence satisfy $\mathrm{ts}\le T$; prior events and retrieved Episodes/Themes must end
before $T$. Skills require every provenance ancestor to resolve before $T$; this ancestry gate
rejects missing or late ancestry. Prompts omit $y(c)$; NDP probes shortcuts and
content but cannot exclude contamination.

\section{Method}
\label{sec:method}

In this section, we introduce the proposed \textbf{TIEM} in detail, including Event-Evidence Hypergraph, Case-based Skill Memory, and Heterogeneous Evidence-Experience Fusion Reasoning.

\subsection{Event-Evidence Hypergraph}
\label{sec:eeh}

EEH organizes timestamped evidence into timestamp-filtered Day, Episode, and Theme
tiers.

\textbf{Day tier.} Each event document yields atomic facts
$h=(\mathcal{S}_h,\rho_h,\mathrm{ts}_h)$ with entity set $\mathcal{S}_h$, fact
$\rho_h$, and source time $\mathrm{ts}_h$. For catalyst $c=(i,T,\kappa)$, $\Phi(c)$ takes
the latest available same-stock source by decision time $T$ as focal text, while
$\kappa$ conditions subsequent semantic retrieval:
\begin{equation}
\Phi(c)=\mathrm{Text}\!\left(\arg\max_{d:\,\mathrm{stock}_d=i,\ \mathrm{ts}_d\le T}\mathrm{ts}_d\right).
\label{eq:eeh-focal}
\end{equation}
Prior Day retrieval uses $q_c=\mathrm{topic}(\kappa)\oplus\mathrm{name}(i)$ to rank
a global semantic candidate pool:
\begin{equation}
\begin{split}
\mathcal G_c={}&\mathrm{Top}_{\xi_1k_1}\{h:\text{ranked by }\mathrm{sim}(h,q_c)\},\\
H_c={}&\mathrm{First}_{k_1}\{h\in\mathcal G_c:s(h)=i,\ \mathrm{ts}_h<T\}\setminus\mathcal F_c,
\end{split}
\label{eq:eeh-concurrent}
\end{equation}
where $s(h)$ gives $h$'s stock, $\mathrm{sim}$ is cosine similarity,
$(\xi_1,k_1)$ set pool multiplier and Day budget, and $\mathcal F_c$ contains
focal document/fact timestamp matches.

\textbf{Episode tier.} EEH greedily partitions each stock $s$ into nonoverlapping
windows $H_w$ bounded by $\Delta_{\mathrm{ep}}$. Windows with at least
$n_{\mathrm{ep}}^{\min}$ hyperedges apply a deterministic child cap to form a
bounded Episode candidate under these constraints:
\begin{equation}
\begin{gathered}
\bar H_w=\mathrm{Cap}_{n_{\mathrm{ep}}^{\max}}(H_w),\;
n_{\mathrm{ep}}^{\min}\le|\bar H_w|\le n_{\mathrm{ep}}^{\max},\\
h_2=\bigl(\textstyle\bigoplus_{h_j\in\bar H_w}h_j,\ s,\rho_2,
\tau_2^{\mathrm{s}},\tau_2^{\mathrm{e}}\bigr),\\
\tau_2^{\mathrm{e}}-\tau_2^{\mathrm{s}}\le\Delta_{\mathrm{ep}},
\end{gathered}
\label{eq:eeh-episode}
\end{equation}
where $\bigoplus$ stores child ids, $\rho_2$ summarizes source-chunk prefixes, and
$[\tau_2^{\mathrm{s}},\tau_2^{\mathrm{e}}]$ is the original extent. Retention
requires at least $v_{\mathrm{ep}}$ of $N_{\mathrm{ep}}$ positive LLM votes with
confidence $\ge\theta_{\mathrm{ep}}$ for retention.

\textbf{Theme tier.} Windows of length $\Delta_{\mathrm{th}}$ and step
$\delta_{\mathrm{th}}$ use HDBSCAN with DBSCAN fallback to cluster Episode-summary
embeddings into candidate Themes before structural filtering as follows:
\begin{equation}
\begin{split}
h_3^{(0)} = \bigl(\textstyle\bigoplus_k h_2^{(k)},\ \{s_k\},\ \rho_3,\ \tau_3^{\mathrm{s}},\ \tau_3^{\mathrm{e}}\bigr),\\
n_{\mathrm{st}}^{\min}\le|\{s_k\}|,\quad |\{h_2^{(k)}\}|\le n_{\mathrm{cl}}^{\max},\\
|\{\mathrm{ind}(s_k):\mathrm{ind}(s_k)\ne\varnothing\}|\ge n_{\mathrm{ind}}^{\min},\\
\tau_3^{\mathrm{s}}=\min_k\tau_{2,k}^{\mathrm{s}},\quad
\tau_3^{\mathrm{e}}=\max_k\tau_{2,k}^{\mathrm{e}},
\end{split}
\label{eq:eeh-theme}
\end{equation}
where $h_3^{(0)}$ is pre-merge; $h_2^{(k)}$, $s_k$, $\rho_3$, and
$[\tau_3^{\mathrm{s}},\tau_3^{\mathrm{e}}]$ are its child Episode, stock, LLM
description, and extent; $\mathrm{ind}(s)$ is the stored industry or $\varnothing$.
Theme construction requires at least $n_{\mathrm{pool}}^{\min}$ stored Episodes per
dataset; $n_{\mathrm{st}}^{\min},n_{\mathrm{ind}}^{\min}$ set stock
and known-industry minima, and $n_{\mathrm{cl}}^{\max}$ caps pre-merge size.
Confidence-qualified, name-similar records merge stocks and child Episodes
across windows; final consolidation may exceed this cap.

\textbf{Timestamp-filtered retrieval.} EEH retrieves the following evidence
package for catalyst $c$:
\begin{equation}
\mathcal{E}(c) = \Phi(c) \oplus U(c) \oplus H_c \oplus \mathrm{Top}_{k_2}^{\mathrm{ep}}(c) \oplus \mathrm{Top}_{k_3}^{\mathrm{th}}(c),
\label{eq:eeh-c2f}
\end{equation}
where $U(c)$ contains up to $n_U$ same-stock decision-day facts available by $T$;
$k_2,k_3$ budget same-stock Episodes and stock-containing Themes; $H_c$ and
retrieved Episode/Theme records precede $T$, while $\Phi(c)$ may
occur at $T$ under the focal gate.

\textbf{Proposition 1.} \textit{EEH can improve forecasting through
timestamp-gated multi-tier evidence retrieval under explicit temporal
constraints.}\vspace{-1.66mm}
\begin{proof}
We provide results in Section~\ref{sec:exp_ablation} and
theoretical proofs in Appendix~\ref{proof1}.
\end{proof}

\subsection{Case-based Skill Memory}
\label{sec:csm}

CSM stores each outcome-revealed skill in source-tagged vector memory as
\begin{equation}
\sigma=(n,\psi,D,a,T_\sigma,s,t,p,e,A),
\label{eq:csm-schema}
\end{equation}
where $n,\psi,D,a,T_\sigma,s,t$ are name,
IF--THEN rule, domain, EMA advantage, tags, stability, and representative
resolve time, with
$s\in\{\textsc{stable},\textsc{unstable},\textsc{miss},\textsc{unevaluable}\}$.
$p$ stores one source and at most $n_p$ ids; $A$ stores complete ancestry with
resolve times; normalized $e\in\mathbb{R}^d$ embeds $n\oplus\psi$; and
$\Sigma=\{\sigma_m\}_{m=1}^M$ is the $M$-skill base for updates and retrieval.

\textbf{Lifecycle.} Once $c$ resolves, an LLM distils $\sigma_{\mathrm{new}}$ with
$e_{\mathrm{new}}$. For nonempty $\Sigma_D=\{\sigma\in\Sigma:\sigma.D=D\}$, let
$\sigma^*=\arg\max_{\sigma'\in\Sigma_D}\cos(e_{\mathrm{new}},e_{\sigma'})$;
EVOKE then applies the following update:
\begin{equation}
\begin{split}
\textsc{Evoke}\ \sigma_{\mathrm{new}}:\quad
&\text{if }\Sigma_D\!\ne\!\varnothing,\ \cos(e_{\mathrm{new}},e_{\sigma^*})\ge\tau,\\[-1mm]
&\text{auto-MERGE }\sigma_{\mathrm{new}}\!\to\!\sigma^*;\\[-1mm]
&\text{else }\Sigma\leftarrow\Sigma\cup\{\sigma_{\mathrm{new}}\},
\end{split}
\label{eq:csm-evoke}
\end{equation}
where $\tau$ is the similarity threshold. Auto-MERGE updates
$\sigma^*.a\leftarrow\lambda_a\sigma^*.a+(1-\lambda_a)a_{\mathrm{new}}$ using
retention $\lambda_a$ and candidate advantage $a_{\mathrm{new}}$, unions tags,
capped source ids, and ancestry, and resets $s$ to \textsc{unevaluable}. Let
$z_i=\sigma_i.z$ for $z\in\{\psi,a,e,s\}$, $T_i=\sigma_i.T_\sigma$, and
$u=\textsc{unevaluable}$; the explicit lifecycle operations are then defined as follows:
\begin{equation}
\begin{gathered}
\textsc{Revise}\ \sigma:\quad
a\leftarrow\mathrm{clip}(a',a_{\min},a_{\max}),\,
\psi\leftarrow\psi',\\
\textsc{Merge}\ \sigma_j\!\to\!\sigma_i:\quad
\psi_i\leftarrow\psi_i\oplus\psi_j,\\
a_i\leftarrow\lambda_m a_i+(1-\lambda_m)a_j,\\[-1mm]
T_i\leftarrow T_i\cup T_j,\quad A_i\leftarrow A_i\cup A_j,\\
e_i\leftarrow\mathrm{Embed}(n_i\oplus\psi_i),\;
s_i\leftarrow u,\\[-1mm]
\Sigma\leftarrow\Sigma\setminus\{\sigma_j\},\\
\textsc{Retire}\ \sigma:\quad
\Sigma\leftarrow\Sigma\setminus\{\sigma\}.
\end{gathered}
\label{eq:csm-ops}
\end{equation}
where $a',\psi'$ are updates; $[a_{\min},a_{\max}],L_\psi,\lambda_m$ set advantage
bounds, description cap, and merge weight. Both merges retain target $t$ as
representative metadata and propagate $A$ as the temporal authority; explicit
MERGE may drop capped source ids from $p$ but preserves their ancestry in $A$.

\textbf{Optional stability diagnosis.} A subset of catalysts is randomly sampled
from $\mathcal A_{\rm train}$. Let $N_A,N_M,N_S$ count \textsc{Apply},
\textsc{Misapply}, and \textsc{Skip}; the resulting statistics and states follow:
\begin{equation}
\begin{gathered}
K_{\mathrm{eff}}=\min(K_{\mathrm{stab}},|\mathcal A_{\rm train}|),\;
n_{\mathrm{rel}}=N_A+N_M,\\
r(\sigma)=N_A/n_{\mathrm{rel}},\quad q_{\mathrm{skip}}=N_S/K_{\mathrm{eff}},\\
s=\begin{cases}
\textsc{unevaluable},&\substack{K_{\mathrm{eff}}<K_{\min}\ \vee\ n_{\mathrm{rel}}=0\\
\vee\ q_{\mathrm{skip}}\ge\eta},\\
\textsc{stable},&r(\sigma)\ge\theta_h,\\
\textsc{miss},&r(\sigma)\le\theta_l,\\
\textsc{unstable},&\text{otherwise}.
\end{cases}
\end{gathered}
\label{eq:csm-stability}
\end{equation}
where $K_{\mathrm{stab}},K_{\min}$ set the sample cap and minimum size, $\eta$ the
skip limit, and $\theta_l,\theta_h$ the miss/stable thresholds. \textsc{Miss}
skills retire; diagnosis uses only catalyst id, stock, type, and outcome,
without event text or internal reasoning traces.

\textbf{Retrieval and optional skill-conditioned Multi-Path Reasoning (SCMR).}
For $c$, $D_c,T_c,e_c$ denote its domain, decision time, and embedding of the
type topic plus the first $L_q$ focal-event characters used to retrieve candidate skills:
\begin{equation}
\begin{split}
b_c(\sigma)={}&\cos(e_c,e_\sigma)+\gamma\mathbf{1}[\sigma.D=D_c],\\
q_c(\sigma)={}&b_c(\sigma)\left(1+\frac{\sigma.a+1}{2}\right),\\
\bar S_k(c)={}&\mathrm{Top}\text{-}k_{\sigma\in\Sigma:\max_{u\in\sigma.A}u.t<T_c}\ q_c(\sigma),\\
S_k(c)={}&\{\sigma\in\bar S_k(c):\cos(e_c,e_\sigma)\ge\mu\},
\end{split}
\label{eq:csm-topk}
\end{equation}
where $b_c,q_c$ are base and advantage-scaled scores, $u.t$ is an ancestor's
resolve time, $\gamma$ is the domain boost, and $\mathbf{1}[\cdot]$ the indicator. $\mu$ forms $S_k(c)$;
optional weighting uses $q_c(\sigma)\alpha(\sigma.s)$. SCMR takes
$m=\min(m_{\max},|S_k(c)|)$ skills as $S_m(c)$ under cap $m_{\max}$; each skill yields a path hypothesis:
\begin{equation}
\begin{split}
\pi_\sigma={}&(\varphi_\sigma,V_\sigma,d_\sigma),\\
V_\sigma\subseteq{}&\{\textsc{focal},\!\textsc{prior-event},
\!\textsc{episode},\!\textsc{theme}\},
\end{split}
\label{eq:csm-hyp}
\end{equation}
where $\varphi_\sigma,V_\sigma$, and
$d_\sigma\in\{\textsc{pos},\textsc{neg}\}$ are the IF--THEN predicate,
evidence types, and direction. The resulting skill-conditioned path call returns:
\begin{equation}
\begin{split}
(\hat{y}_\sigma, w_\sigma, r_\sigma) = f_{\mathrm{LLM}}(\sigma, \pi_\sigma, E_\sigma(c), x_c^{\mathrm{evt}}),\\
\hat{y}_\sigma\!\in\!\{\textsc{pos}, \textsc{neg}, \textsc{abstain}\},
\end{split}
\label{eq:csm-path}
\end{equation}
where $f_{\mathrm{LLM}}$ is inference, $E_\sigma(c)$ the selected evidence,
$x_c^{\mathrm{evt}}$ focal text, $w_\sigma\in[0,1]$ clamped support, and $r_\sigma$
the rationale. Valid non-abstaining paths form $\mathcal V(c)\subseteq S_m(c)$
and jointly produce:
\begin{equation}
\hat{y}(c) = \arg\max_{y\in\{+,-\}}\!\!\!\!\sum_{\substack{\sigma\in\mathcal V(c)\\ \hat{y}_\sigma = y}}\!\!\!\!\alpha(\sigma.s)\!\cdot\!w_\sigma,
\label{eq:csm-converge}
\end{equation}
where $y$ is a direction and $(\alpha_{\mathrm{st}},\alpha_{\mathrm{ue}},
\alpha_{\mathrm{un}},\alpha_{\mathrm{mi}})$ weight \textsc{stable},
\textsc{unevaluable}, \textsc{unstable}, and \textsc{miss}. Ties use a
deterministic catalyst-seeded draw, while empty $\mathcal V(c)$ invokes HEFR.

\textbf{Proposition 2.} \textit{CSM can improve forecasting through
outcome-informed cross-case skill reuse.}
\vspace{-1.6mm}
\begin{proof}
We provide results in Section~\ref{sec:exp_ablation} and
theoretical proofs in Appendix~\ref{proof2}.
\end{proof}

\subsection{Heterogeneous Evidence-Experience Fusion Reasoning}
\label{sec:hefr}

HEFR packs evidence and retrieved CSM skills into a unified context:
\begin{equation}
\begin{split}
C(c)={}&\langle x_c,F(c),U(c),E(c),\\
&R(c),P(c),\mathrm{Exp}(c)\rangle,
\end{split}
\label{eq:hefr-context}
\end{equation}
where $x_c$ is the header; $F,U,E,R,P$ denote top-$k_F$ focal, decision-day concurrent,
same-stock Episode, stock-containing Theme, and prior same-stock Day streams;
$\mathrm{Exp}$ has up to $k$ unrestricted skills. Records in $F,U$ satisfy $\mathrm{ts}\le T_c$,
$P,E,R$ precede $T_c$; all $\mathrm{Exp}$ ancestors resolve before $T_c$.

\textbf{Context organization.} For $\mathcal A=(F,\mathrm{Exp},E,R,P)$,
normalized shares $\boldsymbol{\omega}=(\omega_X)_{X\in\mathcal A}$,
$\sum_X\omega_X=1$, define limits $b_X$ within shared context capacity
$B_{\mathrm{char}}$, where $\sum_Xb_X\le B_{\mathrm{char}}$, and set
$\widetilde X(c)=\mathrm{Trunc}_{b_X}(X(c))$. We serialize the context as
$\widetilde C(c)=\langle x_c,\widetilde F(c),U(c),\widetilde E(c),
\widetilde R(c),\widetilde P(c),\widetilde{\mathrm{Exp}}(c)\rangle$; $U(c)$ is
retained separately outside the shared capacity constraint and remains untruncated.

\textbf{Joint inference.} Template $\Pi$ places $\widetilde C(c)$ in named
segments for one joint cross-stream call:
\begin{equation}
(\hat y(c),r(c))=f_{\mathrm{LLM}}(\Pi(\widetilde C(c))),
\label{eq:hefr-fusion}
\end{equation}
where $\hat y(c)\in\{\textsc{pos},\textsc{neg}\}$ is the direction and $r(c)$ a
citation-optional rationale; successful SCMR replaces this call, otherwise HEFR
handles zero retrieved skills, invalid paths, or inference errors.

\textbf{Proposition 3.} \textit{HEFR can improve forecasting through
single-call joint fusion of heterogeneous evidence and experience streams.}
\vspace{-1.6mm}
\begin{proof}
We provide results in Section~\ref{sec:exp_ablation} and
theoretical proofs in Appendix~\ref{proof3}.
\end{proof}

\definecolor{grpDirect}{rgb}{0.9608,0.9725,0.9804}
\definecolor{grpRetrieval}{rgb}{0.9000,0.9725,0.9843}
\definecolor{grpPosthoc}{rgb}{0.9098,0.9569,0.9882}
\definecolor{grpOurs}{rgb}{0.9608,0.9098,0.9686}
\begin{table*}[t]
\caption{
Main results of \textbf{\textsc{TIEM}} and the baselines across five event-driven forecasting benchmarks, reported on two backbone LLMs with \textbf{best} in bold. Acc and F1 are in \%; MCC is a coefficient; and higher values are better.}

\centering
\fontsize{9pt}{8pt}\selectfont
\setlength{\tabcolsep}{0.6mm}{

\begin{tabular}{l|ccc|ccc|ccc|ccc|ccc}
\toprule
\multirow{2.5}{*}{\textbf{Method}} & \multicolumn{3}{c}{\textbf{Astock}} & \multicolumn{3}{c}{\textbf{FinPURE}} & \multicolumn{3}{c}{\textbf{CMIN-US}} & \multicolumn{3}{c}{\textbf{EDT}} & \multicolumn{3}{c}{\textbf{CSMD}} \\
\cmidrule(lr){2-4} \cmidrule(lr){5-7} \cmidrule(lr){8-10} \cmidrule(lr){11-13} \cmidrule(lr){14-16}
& \textbf{Acc$\uparrow$} & \textbf{MCC$\uparrow$} & \textbf{F1$\uparrow$} & \textbf{Acc$\uparrow$} & \textbf{MCC$\uparrow$} & \textbf{F1$\uparrow$} & \textbf{Acc$\uparrow$} & \textbf{MCC$\uparrow$} & \textbf{F1$\uparrow$} & \textbf{Acc$\uparrow$} & \textbf{MCC$\uparrow$} & \textbf{F1$\uparrow$} & \textbf{Acc$\uparrow$} & \textbf{MCC$\uparrow$} & \textbf{F1$\uparrow$} \\
\midrule
\multicolumn{16}{c}{\textbf{\textit{DeepSeek-V4-Flash}}} \\
\midrule
\rowcolor{grpDirect} LLM Zero-shot & 53.12 & 0.06 & 42.98 & 53.12 & 0.13 & 42.03 & 48.44 & -0.03 & 37.27 & 50.00 & 0.00 & 39.17 & 53.12 & 0.12 & 44.68 \\
\rowcolor{grpDirect} CoT & 53.91 & 0.11 & 41.47 & 53.12 & 0.18 & 39.92 & 49.22 & 0.00 & 32.98 & 50.00 & 0.00 & 34.67 & 51.56 & 0.15 & 37.92 \\
\rowcolor{grpRetrieval} MemGPT & 57.81 & 0.16 & 57.77 & 55.47 & 0.12 & 53.70 & 50.00 & 0.00 & 49.80 & 52.34 & 0.05 & 51.85 & 55.47 & 0.12 & 55.00 \\
\rowcolor{grpRetrieval} Mem0 & 59.38 & 0.19 & 59.34 & 57.03 & 0.14 & 57.03 & 51.56 & 0.05 & 48.33 & 52.34 & 0.05 & 52.27 & 53.91 & 0.08 & 53.84 \\
\rowcolor{grpRetrieval} A-MEM & 60.94 & 0.23 & 60.59 & 57.03 & 0.16 & 53.97 & 50.78 & 0.02 & 49.67 & 52.34 & 0.05 & 52.20 & 53.91 & 0.09 & 48.64 \\
\rowcolor{grpPosthoc} Vanilla RAG & 60.94 & 0.22 & 60.70 & 57.81 & 0.16 & 56.76 & 53.12 & 0.08 & 50.77 & 53.12 & 0.08 & 47.47 & 55.47 & 0.11 & 55.25 \\
\rowcolor{grpPosthoc} HippoRAG & 63.28 & 0.26 & 63.17 & 60.16 & 0.21 & 59.44 & 52.34 & 0.06 & 49.37 & 53.91 & 0.11 & 48.04 & 56.25 & 0.12 & 56.21 \\
\rowcolor{grpPosthoc} GraphRAG & 61.72 & 0.23 & 61.53 & 60.16 & 0.20 & 60.04 & 52.34 & 0.07 & 47.47 & 53.91 & 0.14 & 44.37 & 57.03 & 0.15 & 56.58 \\
\rowcolor{grpPosthoc} LightRAG & 62.50 & 0.25 & 62.35 & 60.16 & 0.21 & 59.86 & 53.12 & 0.08 & 50.40 & 53.91 & 0.12 & 46.72 & 57.81 & 0.19 & 55.36 \\
\rowcolor{grpPosthoc} HyperGraphRAG & 62.50 & 0.25 & 62.27 & 60.94 & 0.22 & 60.70 & 55.47 & 0.13 & 53.06 & 55.47 & 0.13 & 51.41 & 57.81 & 0.16 & 57.44 \\
\rowcolor{grpOurs} \textbf{TIEM (Ours)} & \textbf{65.62} & \textbf{0.31} & \textbf{65.59} & \textbf{69.53} & \textbf{0.43} & \textbf{68.11} & \textbf{62.50} & \textbf{0.26} & \textbf{61.90} & \textbf{59.38} & \textbf{0.23} & \textbf{55.89} & \textbf{64.06} & \textbf{0.30} & \textbf{63.49} \\
\midrule
\multicolumn{16}{c}{\textbf{\textit{GPT-5.4-mini}}} \\
\midrule
\rowcolor{grpDirect} LLM Zero-shot & 50.78 & 0.01 & 50.42 & 50.00 & 0.00 & 40.12 & 50.78 & 0.02 & 50.75 & 52.34 & 0.06 & 48.49 & 49.22 & -0.02 & 48.69 \\
\rowcolor{grpDirect} CoT & 52.34 & 0.20 & 40.56 & 50.00 & 0.00 & 33.33 & 52.34 & 0.05 & 45.62 & 52.34 & 0.10 & 40.56 & 50.78 & 0.00 & 33.68 \\
\rowcolor{grpRetrieval} MemGPT & 55.47 & 0.20 & 49.19 & 52.34 & 0.15 & 38.34 & 53.12 & 0.06 & 50.00 & 53.12 & 0.07 & 51.42 & 51.56 & 0.03 & 43.64 \\
\rowcolor{grpRetrieval} Mem0 & 57.03 & 0.18 & 54.35 & 52.34 & 0.15 & 38.34 & 53.12 & 0.06 & 49.57 & 54.69 & 0.12 & 49.78 & 52.34 & 0.06 & 42.48 \\
\rowcolor{grpRetrieval} A-MEM & 55.47 & 0.13 & 53.98 & 53.91 & 0.11 & 47.41 & 53.12 & 0.06 & 50.00 & 53.91 & 0.08 & 52.37 & 51.56 & 0.04 & 39.05 \\
\rowcolor{grpPosthoc} Vanilla RAG & 58.59 & 0.19 & 57.66 & 57.81 & 0.22 & 52.16 & 55.47 & 0.11 & 55.47 & 53.91 & 0.08 & 53.77 & 53.91 & 0.08 & 50.62 \\
\rowcolor{grpPosthoc} HippoRAG & 58.59 & 0.18 & 58.47 & 56.25 & 0.15 & 52.93 & 55.47 & 0.11 & 55.47 & 55.47 & 0.11 & 55.34 & 53.91 & 0.09 & 47.41 \\
\rowcolor{grpPosthoc} GraphRAG & 60.94 & 0.22 & 60.93 & 58.59 & 0.18 & 57.21 & 54.69 & 0.11 & 52.74 & 55.47 & 0.11 & 55.44 & 53.91 & 0.08 & 53.43 \\
\rowcolor{grpPosthoc} LightRAG & 60.94 & 0.22 & 60.85 & 57.81 & 0.16 & 57.14 & 56.25 & 0.14 & 55.16 & 55.47 & 0.11 & 55.34 & 53.91 & 0.08 & 53.84 \\
\rowcolor{grpPosthoc} HyperGraphRAG & 61.72 & 0.24 & 61.53 & 58.59 & 0.20 & 56.01 & 57.03 & 0.14 & 57.03 & 57.03 & 0.16 & 55.03 & 55.47 & 0.12 & 50.91 \\
\rowcolor{grpOurs} \textbf{TIEM (Ours)} & \textbf{65.62} & \textbf{0.31} & \textbf{65.32} & \textbf{64.84} & \textbf{0.36} & \textbf{61.64} & \textbf{63.28} & \textbf{0.28} & \textbf{62.90} & \textbf{62.50} & \textbf{0.25} & \textbf{62.35} & \textbf{60.16} & \textbf{0.20} & \textbf{59.96} \\
\bottomrule
\end{tabular}}
\label{T1}
\vspace{-0.1mm}
\end{table*}

\section{Experiments}
\label{sec:experiments}

In this section, we present the experimental setup, analysis, and results for the following research questions (RQs): \textbf{RQ1:} Does \textbf{TIEM} outperform current baselines on event-driven forecasting benchmarks? \textbf{RQ2:} What name--date and event-content effects does the per-model Name-Date Probe reveal? \textbf{RQ3:} Does \textbf{TIEM} maintain its lead under the evaluated market and time shifts? \textbf{RQ4:} How much do EEH, CSM, and joint HEFR fusion contribute? \textbf{RQ5:} Does \textbf{TIEM} offer a favorable accuracy--token tradeoff among augmented methods? \textbf{RQ6:} Does \textbf{TIEM} maintain prediction consistency across backbone architectures?

\subsection{Experimental Setup}
\label{sec:exp_setup}

\textbf{Datasets.} We evaluate \textbf{TIEM} on \textbf{Astock}~\cite{astock}, \textbf{FinPURE}, \textbf{CMIN-US}~\cite{cmin}, \textbf{EDT}~\cite{edt}, and \textbf{CSMD}~\cite{csmd}. More details are in Appendix~\ref{sec:dataset-details}.

\textbf{Baselines.} Our baselines cover three families, including direct prompting (\textbf{LLM Zero-shot}, \textbf{CoT}~\cite{cot-prompting}), memory-augmented agents (\textbf{MemGPT}~\cite{memgpt}, \textbf{Mem0}~\cite{mem0}, \textbf{A-MEM}~\cite{a-mem}), and retrieval-augmented generation from flat to structured (\textbf{Vanilla RAG}~\cite{rag-lewis}, \textbf{HippoRAG}~\cite{hipporag}, \textbf{GraphRAG}~\cite{graphrag}, \textbf{LightRAG}~\cite{lightrag}, \textbf{HyperGraphRAG}~\cite{hypergraphrag}). More details are illustrated in Appendix~\ref{sec:baseline-details}.

\textbf{Evaluation Metrics.} We report \textbf{Accuracy} (\textbf{Acc}), \textbf{Matthews Correlation Coefficient} (\textbf{MCC}), \textbf{macro F1}, \textbf{MemBase}, $\boldsymbol{\Delta_{\mathrm{content}}}$, cross-dataset-shift average and worst-case accuracy, \textbf{robustness ratio} $\mathrm{Acc}_{\mathrm{shift}}/\mathrm{Acc}_{\mathrm{ref}}$, \textbf{average final-forecast tokens}, Acc/$\ln$(Tokens), and marginal gain per 1k extra tokens in the five-benchmark aggregate. More details are illustrated in Appendix~\ref{sec:evaluation-details}.

\textbf{Implementation.} We evaluate \textbf{TIEM} and all baselines on two LLM backbones: DeepSeek-V4-Flash and GPT-5.4-mini. All methods use official APIs. \textbf{TIEM} uses text-embedding-3-small for EEH. More details are shown in Appendix~\ref{sec:implementation-details}.

\begin{figure*}[t]
  \centering
  \includegraphics[width=1.0\linewidth]{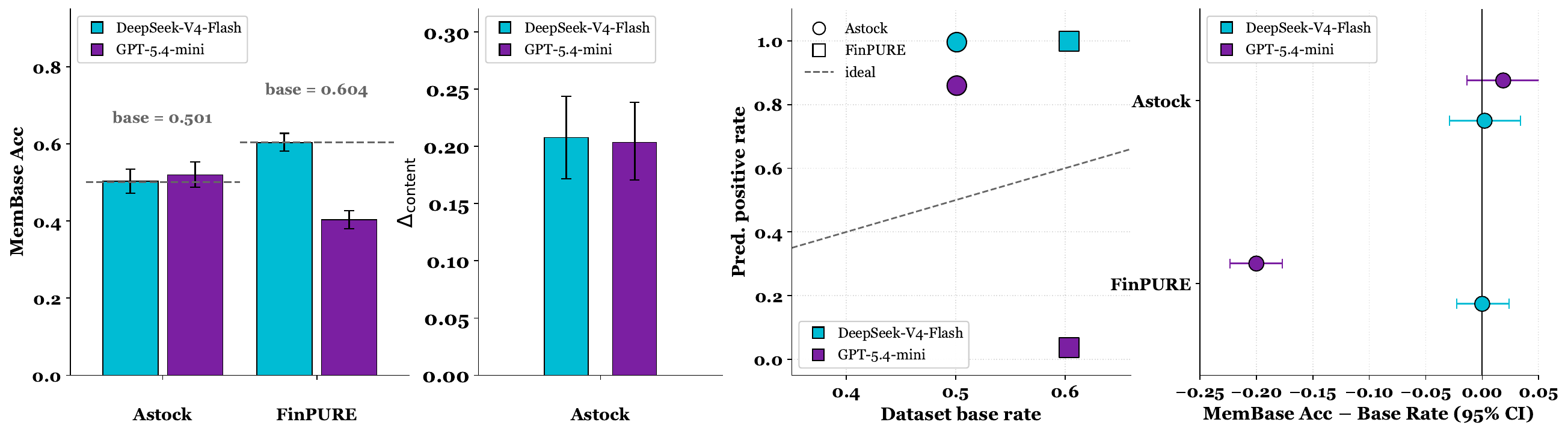}\\
  \makebox[0.27\linewidth]{\small (a) MemBase Acc}\hfill
  \makebox[0.19\linewidth]{\small (b) $\Delta_{\mathrm{content}}$}\hfill
  \makebox[0.26\linewidth]{\small (c) Pred.\ vs.\ base rate}\hfill
  \makebox[0.26\linewidth]{\small (d) Forest plot}
  \vspace{-1mm}
  \caption{Per-model NDP results on independent \textbf{Astock} ($n=968$) and \textbf{FinPURE} ($n=1{,}635$) diagnostic pools under two LLM backbones. Intervals are $95\%$ bootstrap CIs; content gain uses paired bootstrap resampling.}
  \label{fig:rq2_ndp}
\end{figure*}

\begin{figure*}[t]
  \centering
  \includegraphics[width=1.0\linewidth]{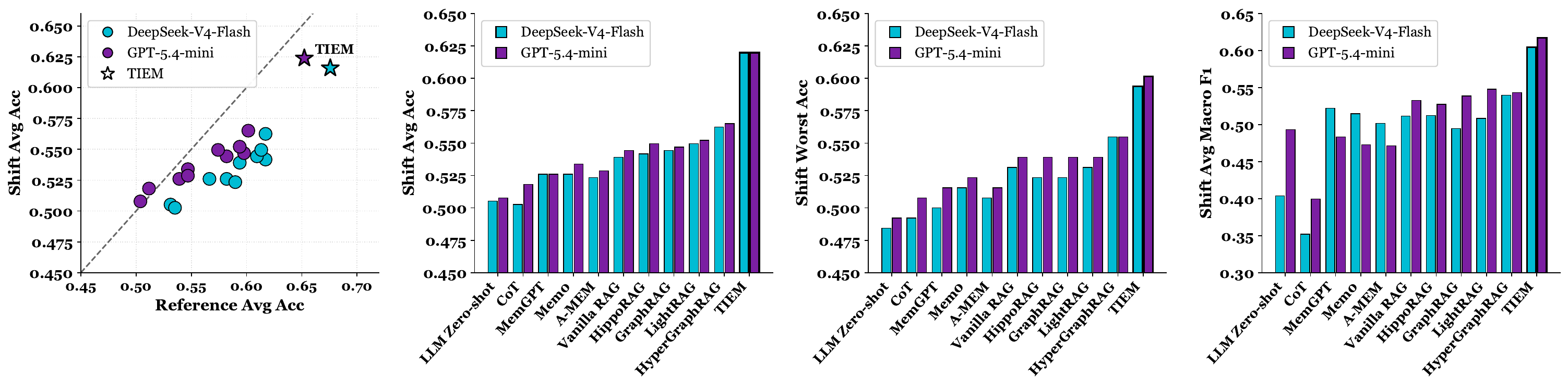}\\
  \makebox[0.25\linewidth]{\small (a) Reference vs.\ shift accuracy}\hfill
  \makebox[0.25\linewidth]{\small (b) Shift-average accuracy}\hfill
  \makebox[0.25\linewidth]{\small (c) Shift worst-case accuracy}\hfill
  \makebox[0.25\linewidth]{\small (d) Shift-average macro F1}
  \vspace{-4.6mm}
  \caption{Per-method performance on cross-dataset \textbf{CMIN-US}, \textbf{EDT}, and \textbf{CSMD} shifts. The reference averages Astock and FinPURE, while shift averages and worst cases are computed over CMIN-US, EDT, and CSMD.}
  \label{fig:rq3_shift}
\end{figure*}

\begin{figure*}[t]
  \centering
  \includegraphics[width=1.0\linewidth]{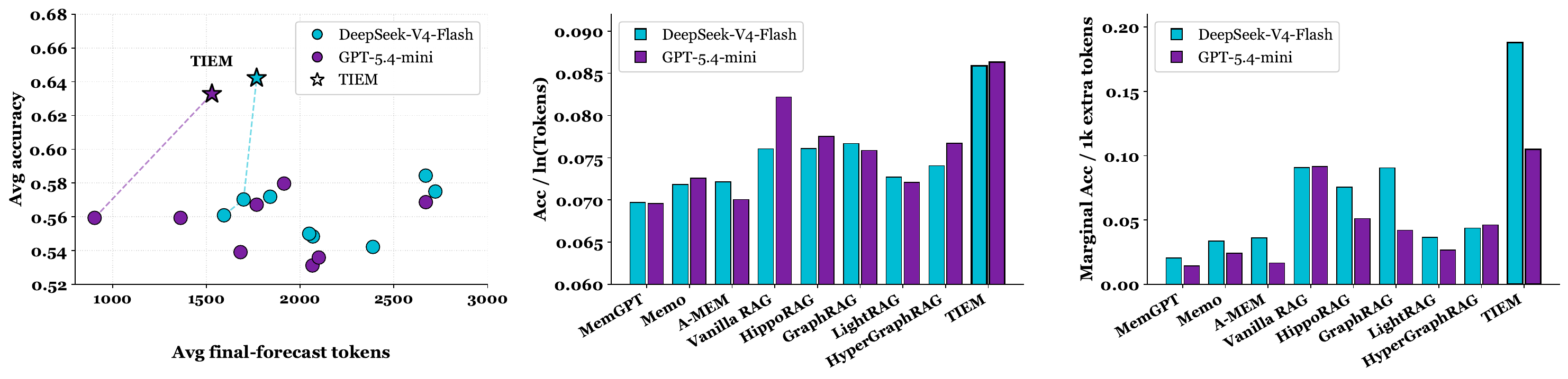}\\
  \makebox[0.33\linewidth]{\small (a) Token-vs-accuracy Pareto scatter}\hfill
  \makebox[0.33\linewidth]{\small (b) Acc / $\ln$(Tokens)}\hfill
  \makebox[0.33\linewidth]{\small (c) Marginal Acc per 1k tokens}
  \vspace{-1mm}
  \caption{Final-forecast accuracy--token tradeoff of \textsc{TIEM} across five benchmarks and two LLM backbones. Dashed lines denote per-backbone Pareto frontiers, with marginal gains measured against the Zero-shot reference.}
  \label{fig:rq5_costeff}
\end{figure*}

\begin{figure*}[t]
  \centering
  \includegraphics[width=1.0\linewidth]{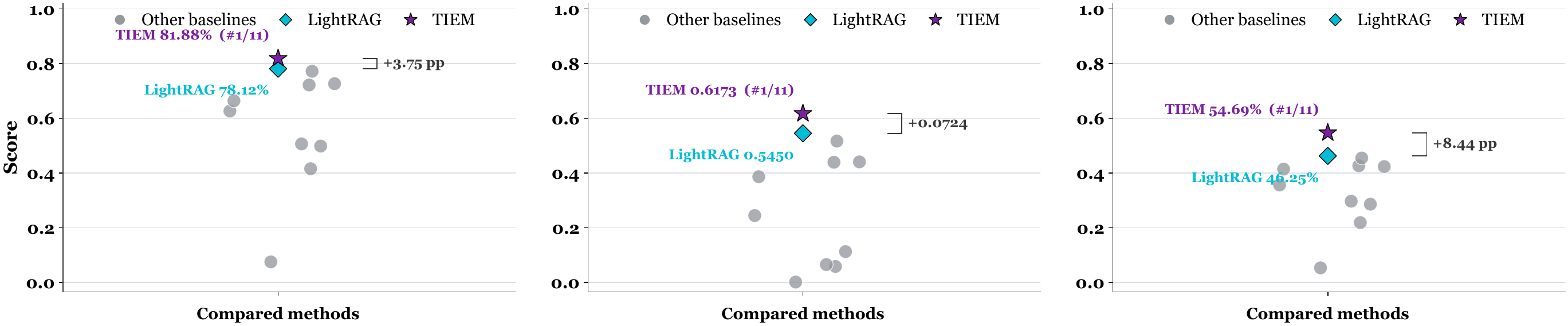}\\
  \makebox[0.33\linewidth]{\small (a) Prediction agreement}\hfill
  \makebox[0.33\linewidth]{\small (b) Cohen's $\kappa$}\hfill
  \makebox[0.33\linewidth]{\small (c) Dual-correct rate}
  \vspace{-1mm}
  \caption{Macro-averaged cross-backbone prediction consistency across evaluation datasets. \textbf{\textsc{TIEM}} ranks first among eleven methods in agreement, Cohen's $\kappa$, and dual-correct rate in the reported comparisons.}
  \label{fig:rq6_macro}
\end{figure*}

\subsection{Main Results (RQ1)}
\label{sec:exp_main}

Table~\ref{T1} compares \textbf{\textsc{TIEM}} with ten baselines on five forecasting benchmarks. The results reveal: \textit{(i)} \textbf{Cross-backbone leadership.} \textbf{\textsc{TIEM}} achieves the strongest average performance on both backbones, showing that its evidence and memory integration transfers across model choices. \textit{(ii)} \textbf{Cross-benchmark leadership.} It leads all three metrics on every benchmark, showing that timestamped evidence and transferable skills remain effective across diverse markets and periods. \textit{(iii)} \textbf{Class-balanced improvement.} Its concurrent gains in MCC and macro F1 extend the accuracy advantage to balanced predictive quality across outcome classes on the two large language model backbones.

\subsection{Name--Date Sensitivity Probe (RQ2)}
\label{sec:exp_ndp}

As shown in Fig.~\ref{fig:rq2_ndp}, the per-model NDP reveals three findings: \textit{(i)} \textbf{Name--date resistance.} MemBase does not exceed the base rate under the reported CIs on either dataset, showing that names and dates alone provide no reliable forecasting advantage. \textit{(ii)} \textbf{Event-content contribution.} On Astock, focal-event text yields positive gains for both backbones beyond identifiers, recent statistics, and background evidence. \textit{(iii)} \textbf{Content-grounded forecasting.} The contrast between weak name--date performance and positive content gains shows that event semantics contribute discriminative information beyond name--date cues for prediction.

\subsection{Cross-Dataset Shift Performance (RQ3)}
\label{sec:exp_shift}

Figure~\ref{fig:rq3_shift} reveals three findings: \textit{(i)} \textbf{Leading shift performance.} \textbf{\textsc{TIEM}} ranks first in average accuracy on both backbones across cross-market and cross-time shifts, showing that its aggregate advantage transfers across model choices and heterogeneous datasets. \textit{(ii)} \textbf{Worst-case strength.} \textbf{\textsc{TIEM}} leads in each method's minimum across the three shift benchmarks, showing that its transfer advantage is sustained across the individual shift datasets rather than concentrated in one result. \textit{(iii)} \textbf{Class-balanced transfer.} \textbf{\textsc{TIEM}} also leads in shift-average macro F1, showing that timestamped evidence and transferable skills preserve class-balanced prediction across market--time shifts.

\begin{table}[t]
\caption{Ablation results on Astock. \textit{w/o} EEH, CSM, and HEFR remove temporal event organization, outcome-informed skill memory, and joint evidence--experience fusion from Full TIEM, respectively.}
\label{tab:rq4_ablation}
\centering
\fontsize{9pt}{8pt}\selectfont
\setlength{\tabcolsep}{1.8pt}
\begin{tabular}{l|ccc|ccc}
\toprule
\multirow{2}{*}{\textbf{Method}} & \multicolumn{3}{c|}{\textbf{DeepSeek-V4-Flash}} & \multicolumn{3}{c}{\textbf{GPT-5.4-mini}} \\
\cmidrule(lr){2-4} \cmidrule(lr){5-7}
& \textbf{Acc$\uparrow$} & \textbf{MCC$\uparrow$} & \textbf{F1$\uparrow$} & \textbf{Acc$\uparrow$} & \textbf{MCC$\uparrow$} & \textbf{F1$\uparrow$} \\
\midrule
\textit{w/o} EEH  & 63.28 & 0.26 & 63.01 & 52.34 & 0.12 & 44.16 \\
\textit{w/o} HEFR & 64.84 & 0.30 & 64.67 & 57.03 & 0.15 & 56.82 \\
\textit{w/o} CSM  & 64.06 & 0.28 & 63.34 & 58.59 & 0.17 & 58.59 \\
\rowcolor{grpOurs} \textbf{TIEM (Full)} & \textbf{65.62} & \textbf{0.31} & \textbf{65.59} & \textbf{65.62} & \textbf{0.31} & \textbf{65.32} \\
\bottomrule
\end{tabular}
\end{table}

\subsection{Ablation Study (RQ4)}
\label{sec:exp_ablation}

Table~\ref{tab:rq4_ablation} compares Full TIEM with three component variants on Astock. The results show: \textit{(i)} \textbf{EEH contribution.} Its gains over \textit{w/o} EEH across metrics and backbones show that temporally organized event evidence helps identify catalyst effects. \textit{(ii)} \textbf{CSM contribution.} Its gains over \textit{w/o} CSM across metrics and backbones show that outcome-informed skills connect current catalysts with reusable experience from resolved cases. \textit{(iii)} \textbf{HEFR contribution.} Its gains over \textit{w/o} HEFR across metrics show that joint evidence--experience synthesis improves directional accuracy and class-balanced prediction by aligning current evidence with transferable skills across both backbones.

\subsection{Accuracy--Token Trade-off (RQ5)}
\label{sec:exp_costeff}

Figure~\ref{fig:rq5_costeff} compares accuracy with final-forecast token use across the benchmarks and backbones. The results show: \textit{(i)} \textbf{Pareto position.} Among displayed methods, \textbf{\textsc{TIEM}} lies on both Pareto frontiers with the highest average accuracy; none improves both token use and accuracy. \textit{(ii)} \textbf{Log-adjusted trade-off.} \textbf{\textsc{TIEM}} leads Acc/$\ln$(Tokens) on both backbones, retaining its accuracy advantage after accounting for final-forecast token use. \textit{(iii)} \textbf{Marginal token trade-off.} Relative to Zero-shot, \textbf{\textsc{TIEM}} yields the largest accuracy gain per 1k additional tokens on both backbones, showing that additional tokens translate effectively into predictive improvement through focused evidence use.

\subsection{Cross-Backbone Consistency (RQ6)}
\label{sec:exp_reliability}

Figure~\ref{fig:rq6_macro} compares cross-backbone agreement, Cohen's $\kappa$, and dual-correct rate. The results show: \textit{(i)} \textbf{Cross-backbone consistency.} \textbf{\textsc{TIEM}} leads all three macro-averaged measures, showing stable predictions from structured evidence and skill fusion. \textit{(ii)} \textbf{Agreement beyond marginals.} Its joint lead in agreement and Cohen's $\kappa$ shows consistency beyond marginal prediction frequencies rather than class prevalence. \textit{(iii)} \textbf{Joint correctness.} Its leading dual-correct rate links agreement with correct outcomes and forecasting quality rather than merely matched predictions across backbones.

\section{Conclusion}
\label{sec:conclusion}

In this work, we propose \textbf{\textsc{TIEM}} for timestamp-gated event-driven catalyst-outcome forecasting. It unifies timestamp-filtered event-evidence retrieval, outcome-informed case-based skill memory, and heterogeneous evidence-experience fusion. In addition, we introduce CIP, FinPURE, and a per-model Name-Date Probe for contamination inspection. Experimental results show benchmark gains, cross-backbone consistency,
and a favorable final-forecast accuracy--token tradeoff. We hope \textbf{\textsc{TIEM}} can provide a new perspective on temporally constrained event-driven financial forecasting.

\bibliography{IEEEfull}

\newpage
\appendix

\section*{Appendix}
\label{sec:appendix}

\section{Theoretical Proof}
\label{sec:theoretical-proof}

\subsection{Proof of Proposition 1}
\label{proof1}
\textbf{Proposition 1.} \textit{EEH can improve forecasting through
timestamp-gated multi-tier evidence retrieval under explicit temporal
constraints.}
\vspace{-1.6mm}
\begin{proof}
Let $c\sim\mathcal P$ be a catalyst drawn from the forecasting
distribution, with decision time $T_c$ and outcome $Y_c\in\{-1,+1\}$.
Let $\mathcal R_{\mathrm{avail}}(T_c)$ contain records actually
available by $T_c$. Write
$E_c^{\mathrm{ep}}$ and $E_c^{\mathrm{th}}$ for the retrieved Episode and
Theme sets. The focal-only and full EEH record collections are
\begin{equation}
\begin{split}
\mathcal Z_0(c)&=\{\Phi(c)\},\\
\mathcal Z_1(c)&=\mathcal Z_0(c)\cup U(c)\cup H_c
\cup E_c^{\mathrm{ep}}\cup E_c^{\mathrm{th}}.
\end{split}
\label{eq:proof-eeh-records}
\end{equation}
With timestamp gates enabled and valid availability metadata, all
records in $\mathcal Z_1(c)$ belong to
$\mathcal R_{\mathrm{avail}}(T_c)$ under compatible time semantics.
Specifically, focal and concurrent records satisfy $\mathrm{ts}\le T_c$;
date-only timestamps use end-of-day semantics. Prior Day records satisfy
$\mathrm{ts}<T_c$, and
Episode and Theme records satisfy $\tau_{\mathrm{end}}<T_c$. Stored
timestamps must denote actual availability. Define the availability indicator and
restricted collection by
\begin{equation}
\begin{split}
I_c(r)&=\mathbf{1}
\{r\in\mathcal R_{\mathrm{avail}}(T_c)\},\\
\mathcal Z_1^{\mathrm{avail}}(c)&=
\{r\in\mathcal Z_1(c):I_c(r)=1\},\\
\sum_{r\in\mathcal Z_1(c)}[1-I_c(r)]&=0,\\
&\Longrightarrow\mathcal Z_1^{\mathrm{avail}}(c)=\mathcal Z_1(c).
\end{split}
\label{eq:proof-eeh-availability}
\end{equation}
Thus, under the stated conditions, the expanded collection is
non-anticipating.

We now give a constructive binary signal-detection proof of the
proposition's existential wording. Index the focal, concurrent, prior Day,
Episode, and Theme evidence components by $j=0,1,2,3,4$, respectively.
Consider a sufficient setting in which the decision-relevant content exposed
by component $j$ admits the scalar statistic
\begin{equation}
X_j=\zeta_jY_c+\varepsilon_j,\qquad
\varepsilon_j\mid Y_c\sim\mathcal N(0,\nu_j^2),
\label{eq:proof-eeh-signal}
\end{equation}
where $\nu_j^2>0$, $\zeta_j\ge0$, and the noise terms are
conditionally independent given $Y_c$. Here $\zeta_j=0$ represents an
uninformative component. This Gaussian construction is an explicit witness for
the existential claim, not a distributional model of arbitrary
natural-language evidence.

For a set of admitted components $\mathcal I$ and
$y\in\{-1,+1\}$, conditional independence gives the joint log-density
\begin{equation}
\begin{split}
\log p(\mathbf X_{\mathcal I}\mid Y_c=y)
&=-\frac12\sum_{j\in\mathcal I}\log(2\pi\nu_j^2)\\
&\quad-\frac12\sum_{j\in\mathcal I}
\frac{(X_j-y\zeta_j)^2}{\nu_j^2}.
\end{split}
\label{eq:proof-eeh-factorization}
\end{equation}
Subtracting these log-densities gives the log-likelihood ratio
\begin{equation}
\begin{split}
\Lambda_{\mathcal I}(\mathbf X)
&=\log\frac{p(\mathbf X_{\mathcal I}\mid Y_c=+1)}
{p(\mathbf X_{\mathcal I}\mid Y_c=-1)}\\
&=2\sum_{j\in\mathcal I}
\frac{\zeta_jX_j}{\nu_j^2}.
\end{split}
\label{eq:proof-eeh-llr}
\end{equation}
Under equal class priors and $0$--$1$ loss, the likelihood-ratio decision is
the sign of
\begin{equation}
S_{\mathcal I}=\sum_{j\in\mathcal I}
\frac{\zeta_jX_j}{\nu_j^2},\qquad
\Gamma_{\mathcal I}=\sum_{j\in\mathcal I}
\frac{\zeta_j^2}{\nu_j^2}.
\label{eq:proof-eeh-statistic}
\end{equation}
The Bayes minimum-error rule, with uniform tie-breaking, is
\begin{equation}
\begin{gathered}
\widehat Y_{\mathcal I}=
\begin{cases}
+1,&S_{\mathcal I}>0,\\
-1,&S_{\mathcal I}<0,
\end{cases}\\
\Pr(\widehat Y_{\mathcal I}=+1\mid S_{\mathcal I}=0)=\tfrac12,\\
\Pr(\widehat Y_{\mathcal I}=-1\mid S_{\mathcal I}=0)=\tfrac12.
\end{gathered}
\label{eq:proof-eeh-bayes}
\end{equation}
Substituting the signal model gives
\begin{equation}
S_{\mathcal I}\mid Y_c=y
\sim\mathcal N(y\Gamma_{\mathcal I},\Gamma_{\mathcal I}),
\qquad y\in\{-1,+1\}.
\label{eq:proof-eeh-distribution}
\end{equation}
Let $Q(x)=\Pr(Z>x)$ for $Z\sim\mathcal N(0,1)$. The resulting
classification error is therefore
\begin{equation}
\begin{split}
P_{\mathrm e}(\mathcal I)
&=\Pr(Y_cS_{\mathcal I}<0)\\
&\quad+\tfrac12\Pr(S_{\mathcal I}=0)\\
&=Q(\sqrt{\Gamma_{\mathcal I}}).
\end{split}
\label{eq:proof-eeh-error}
\end{equation}

For focal-only retrieval, let $\mathcal I_0=\{0\}$; for EEH, let
$\mathcal I_1=\{0,1,2,3,4\}$. Their effective detection strengths obey
\begin{equation}
\Gamma_{\mathcal I_1}
=\Gamma_{\mathcal I_0}
+\sum_{j=1}^{4}\frac{\zeta_j^2}{\nu_j^2}
\ge\Gamma_{\mathcal I_0}.
\label{eq:proof-eeh-strength}
\end{equation}
Because $Q$ is strictly decreasing, any admitted additional component with
$\zeta_j>0$ gives
\begin{equation}
P_{\mathrm e}(\mathcal I_1)
<P_{\mathrm e}(\mathcal I_0).
\label{eq:proof-eeh-strict}
\end{equation}
Hence a timestamp-admissible component carrying a finite-noise, outcome-relevant signal
strictly improves the binary forecast in this realizable setting whenever
the predictor realizes the likelihood-ratio decision.
In summary, the time gates establish admissibility, while the strict error
inequality gives a realizable accuracy gain. Thus EEH can improve forecasting
through timestamp-gated multi-tier retrieval under these sufficient conditions;
this result is not a guarantee for arbitrary evidence or predictors.
\end{proof}

\subsection{Proof of Proposition 2}
\label{proof2}
\textbf{Proposition 2.} \textit{CSM can improve forecasting through
outcome-informed cross-case skill reuse.}
\vspace{-1.6mm}
\begin{proof}
Use the same catalyst notation $c\sim\mathcal P$, $T_c$, and $Y_c$.
Let $\Sigma_{T_c}$ contain only skills whose contributing source outcomes
are all resolved before $T_c$. This provenance condition is enforced by the
complete-ancestry gate, which propagates contributing sources through every
merge and rejects missing, invalid, or late ancestry. It therefore holds for
the chronologically frozen memory used at evaluation, independently of the
representative time retained as auxiliary metadata.

Consider an EMA-updated skill $\sigma$. Let $g_j\in[-1,1]$ be the
outcome-derived feedback proxy at update $j$, let fixed
$a_0\in[-1,1]$, and use
the implemented retention factor $0\le\lambda_a<1$. Let
$\mathcal Q_{\mathrm{store}}$ denote the implementation's storage quantizer. For
this existential witness, choose a feedback process and initialization whose
pre-quantized updates are fixed points:
\begin{equation}
\begin{split}
a_j^{\mathrm{pre}}&=\lambda_a a_{j-1}+(1-\lambda_a)g_j,\\
a_j&=\mathcal Q_{\mathrm{store}}(a_j^{\mathrm{pre}})=a_j^{\mathrm{pre}}.
\end{split}
\label{eq:proof-csm-quantizer}
\end{equation}
The stored update then has the exact recursion and expansion
\begin{equation}
\begin{split}
a_j&=\lambda_a a_{j-1}+(1-\lambda_a)g_j,\\
a_n&=\lambda_a^na_0+(1-\lambda_a)
\sum_{j=1}^{n}\lambda_a^{n-j}g_j.
\end{split}
\label{eq:proof-csm-recursion}
\end{equation}
The coefficients in the second line are nonnegative and sum to one, so
$|a_n|\le1$. In the zero-state form, the corresponding discrete-time
input-output transfer function is
\begin{equation}
\mathscr H_\sigma(z)
=\frac{A(z)}{G(z)}
=\frac{1-\lambda_a}{1-\lambda_a z^{-1}}.
\label{eq:proof-csm-transfer}
\end{equation}
Its only pole is $\lambda_a$, which lies strictly inside the unit circle.
Moreover, its impulse weights
\begin{equation}
\iota_\ell=(1-\lambda_a)\lambda_a^{\ell},\qquad
\sum_{\ell=0}^{\infty}|\iota_\ell|=1,
\label{eq:proof-csm-impulse}
\end{equation}
are absolutely summable. For the zero-state response
$a^{\mathrm{zs}}=\iota*g$, the convolution bound gives
\begin{equation}
\|a^{\mathrm{zs}}\|_\infty
\le\|\iota\|_1\|g\|_\infty
=\|g\|_\infty\le1.
\label{eq:proof-csm-bibo}
\end{equation}
This establishes bounded-input--bounded-output stability. The homogeneous
response also satisfies $|\lambda_a^na_0|\le1$, so bounded outcome feedback
cannot produce an unbounded advantage state, and the contribution of an update
$\ell$ steps in the past decays as $\lambda_a^{\ell}$.

We next quantify concentration above a retrieval threshold. Suppose the
$g_j$ are independent with common mean $\mu_\sigma>0$. Define
\begin{equation}
\begin{split}
\bar a_n&=\mathbb E[a_n]
=\mu_\sigma+\lambda_a^n(a_0-\mu_\sigma),\\
\Omega_n&=\sum_{j=1}^{n}
\bigl((1-\lambda_a)\lambda_a^{n-j}\bigr)^2\\
&=\frac{1-\lambda_a}{1+\lambda_a}
(1-\lambda_a^{2n}).
\end{split}
\label{eq:proof-csm-meanweight}
\end{equation}
For any $\vartheta<\bar a_n$, the weighted Hoeffding bound for
$g_j\in[-1,1]$ gives
\begin{equation}
\Pr(a_n\le\vartheta)
\le\exp\left(
-\frac{(\bar a_n-\vartheta)^2}{2\Omega_n}\right).
\label{eq:proof-csm-hoeffding}
\end{equation}
This result concerns an outcome-feedback proxy, not the counterfactual
utility of an individual skill.

For the reported configuration, with optional stability reweighting disabled,
the code ranks an eligible skill by
\begin{equation}
\begin{split}
b_c(\sigma)&=\cos(e_c,e_\sigma)
+\gamma\mathbf 1[\sigma.D=D_c],\\
q_c(\sigma)&=b_c(\sigma)\frac{3+a_\sigma}{2}.
\end{split}
\label{eq:proof-csm-score}
\end{equation}
Whenever $b_c(\sigma)>0$,
$\partial q_c(\sigma)/\partial a_\sigma=b_c(\sigma)/2>0$. For this witness,
take at least $k$ other time-eligible candidates whose scores are deterministic
constants independent of $g_{1:n}$, and let $q_c^{\mathrm{cut}}$ be their
$k$-th largest score. Define the required advantage
\begin{equation}
a_c^{\mathrm{req}}=
\frac{2q_c^{\mathrm{cut}}}{b_c(\sigma)}-3.
\label{eq:proof-csm-margin}
\end{equation}
After $n$ EMA updates, $\sigma$ enters the ranked set whenever it passes
the cosine gate and $a_\sigma=a_n>a_c^{\mathrm{req}}$. If
$a_c^{\mathrm{req}}<\bar a_n$, applying the weighted bound at this threshold
gives
\begin{equation}
\begin{split}
\Pr\bigl(q_c(\sigma)\le q_c^{\mathrm{cut}}\bigr)
&=\Pr(a_n\le a_c^{\mathrm{req}})\\
&\le\exp\left(
-\frac{(\bar a_n-a_c^{\mathrm{req}})^2}
{2\Omega_n}\right).
\end{split}
\label{eq:proof-csm-retrieval-bound}
\end{equation}
This connects resolved outcomes to a bounded, advantage-sensitive retrieval
margin.

Finally, consider a positive-probability family $\mathcal W_\sigma$ of
cross-case catalysts for which $\sigma$ passes the time, cosine, and
realized score-margin gates, and its IF--THEN rule is applicable and returns
$Y_c$ exactly. Suppose the EEH-only decision
$d_{\mathrm E}$ is conditionally uninformative on $\mathcal W_\sigma$,
so its error is $1/2$, the CSM-conditioned decision $d_{\mathrm C}$ follows
the matched rule there, and both decisions agree outside
$\mathcal W_\sigma$. Then
\begin{equation}
\mathrm{Acc}(d_{\mathrm C})-\mathrm{Acc}(d_{\mathrm E})
=\frac{1}{2}\Pr(c\in\mathcal W_\sigma)>0.
\label{eq:proof-csm-witness}
\end{equation}
This is a realizable strict-gain witness: resolved outcome feedback yields a
bounded proxy and the stated retrieval-margin bound, the implemented score
retrieves the applicable rule, and cross-case reuse corrects otherwise
ambiguous cases. In conclusion, CSM can improve
forecasting through outcome-informed cross-case skill reuse under the stated
conditions. This establishes predictive reuse, not causal identification of
an LLM-distilled rule, and does not assert that every skill transfers.
\end{proof}

\subsection{Proof of Proposition 3}
\label{proof3}
\textbf{Proposition 3.} \textit{HEFR can improve forecasting through
single-call joint fusion of heterogeneous evidence and experience streams.}
\vspace{-1.6mm}
\begin{proof}
Let $Y_c\in\{-1,+1\}$ be the forecasting target. Index the five information streams in the implemented order
$\mathcal A=(F,\mathrm{Exp},E,R,P)$ by $i=1,\ldots,5$. Let $Z_i$ denote
the observation supplied by stream $i$, and define the optimal single-stream
and joint Bayes risks as
\begin{equation}
\begin{split}
R_i^*&=\inf_{\delta_i}\Pr\!\left(\delta_i(Z_i)\ne Y_c\right),\\
R_{\mathrm J}^*&=\inf_{\delta}\Pr\!\left(\delta(Z_1,\ldots,Z_5)\ne Y_c\right).
\end{split}
\label{eq:proof-hefr-bayes}
\end{equation}
Every single-stream rule is also a joint rule that ignores the other
observations. Hence the joint decision class contains each single-stream
class, and
\begin{equation}
R_{\mathrm J}^*\le\min_{1\le i\le5}R_i^*.
\label{eq:proof-hefr-noninferior}
\end{equation}
We next give a standard information-theoretic construction in which this
inequality is strict. Suppose each stream exposes a noisy directional cue
\begin{equation}
\begin{split}
\chi_i&=Y_cN_i,\\
\Pr(N_i=-1\mid Y_c=y)&=\epsilon_{\mathrm{ch}},\\
\Pr(N_i=+1\mid Y_c=y)&=1-\epsilon_{\mathrm{ch}},
\end{split}
\label{eq:proof-hefr-channel}
\end{equation}
for every $y\in\{-1,+1\}$, where $0<\epsilon_{\mathrm{ch}}<1/2$ and
$N_1,\ldots,N_5$ are conditionally independent given $Y_c$. This is the
classical five-use binary symmetric-channel setting. A decision based on one
cue has error $\epsilon_{\mathrm{ch}}$. The maximum-likelihood decoder for the
associated length-five repetition code is the odd-majority rule. A joint
decision function with access to all five streams can represent this decoder as
\begin{equation}
d_{\mathrm J}
=\operatorname{sign}\left(\sum_{i=1}^{5}\chi_i\right).
\label{eq:proof-hefr-majority}
\end{equation}
It is wrong only when at least three cues are wrong. Its error is
characterized by the count
\begin{equation}
\begin{split}
K&=\sum_{i=1}^{5}\mathbf{1}\{N_i=-1\},\\
K\mid Y_c=y&\sim\operatorname{Binomial}(5,\epsilon_{\mathrm{ch}}),\\
\{d_{\mathrm J}\ne Y_c\}&=\{K\ge3\}.
\end{split}
\label{eq:proof-hefr-error-count}
\end{equation}
Thus
\begin{equation}
\begin{split}
P_{\mathrm e}^{(5)}
&=\sum_{k=3}^{5}\binom{5}{k}
\epsilon_{\mathrm{ch}}^k(1-\epsilon_{\mathrm{ch}})^{5-k}\\
&=10\epsilon_{\mathrm{ch}}^3-15\epsilon_{\mathrm{ch}}^4
+6\epsilon_{\mathrm{ch}}^5.
\end{split}
\label{eq:proof-hefr-majority-error}
\end{equation}
Direct factorization yields
\begin{equation}
\begin{split}
\epsilon_{\mathrm{ch}}-P_{\mathrm e}^{(5)}
&=\epsilon_{\mathrm{ch}}(1-\epsilon_{\mathrm{ch}})
(1-2\epsilon_{\mathrm{ch}})\\
&\quad\cdot(1+3\epsilon_{\mathrm{ch}}
-3\epsilon_{\mathrm{ch}}^2)>0,
\end{split}
\label{eq:proof-hefr-strict}
\end{equation}
because every factor is positive for $0<\epsilon_{\mathrm{ch}}<1/2$. Hence
the joint decoder is strictly more reliable than any single cue. In this
construction, a single-stream decision has error $\epsilon_{\mathrm{ch}}$,
whereas the joint decision has error $P_{\mathrm e}^{(5)}<\epsilon_{\mathrm{ch}}$.

HEFR exposes the focal, experience, episodic, thematic, and prior-event
streams to one inference function in a named context. Its joint decision
class can therefore represent the decoder above. The construction provides
an explicit witness that joint fusion can reduce forecasting error when
heterogeneous streams contain complementary cues. It is an existence
argument for the fusion mechanism; it neither assumes that real streams are
independent and identically distributed nor asserts that every LLM call
realizes the optimal decoder.
\end{proof}

\section{Prompts Used in \textsc{TIEM}}
\label{sec:prompts}

\begin{figure*}[p]
    \centering
    \includegraphics[width=1.0\textwidth]{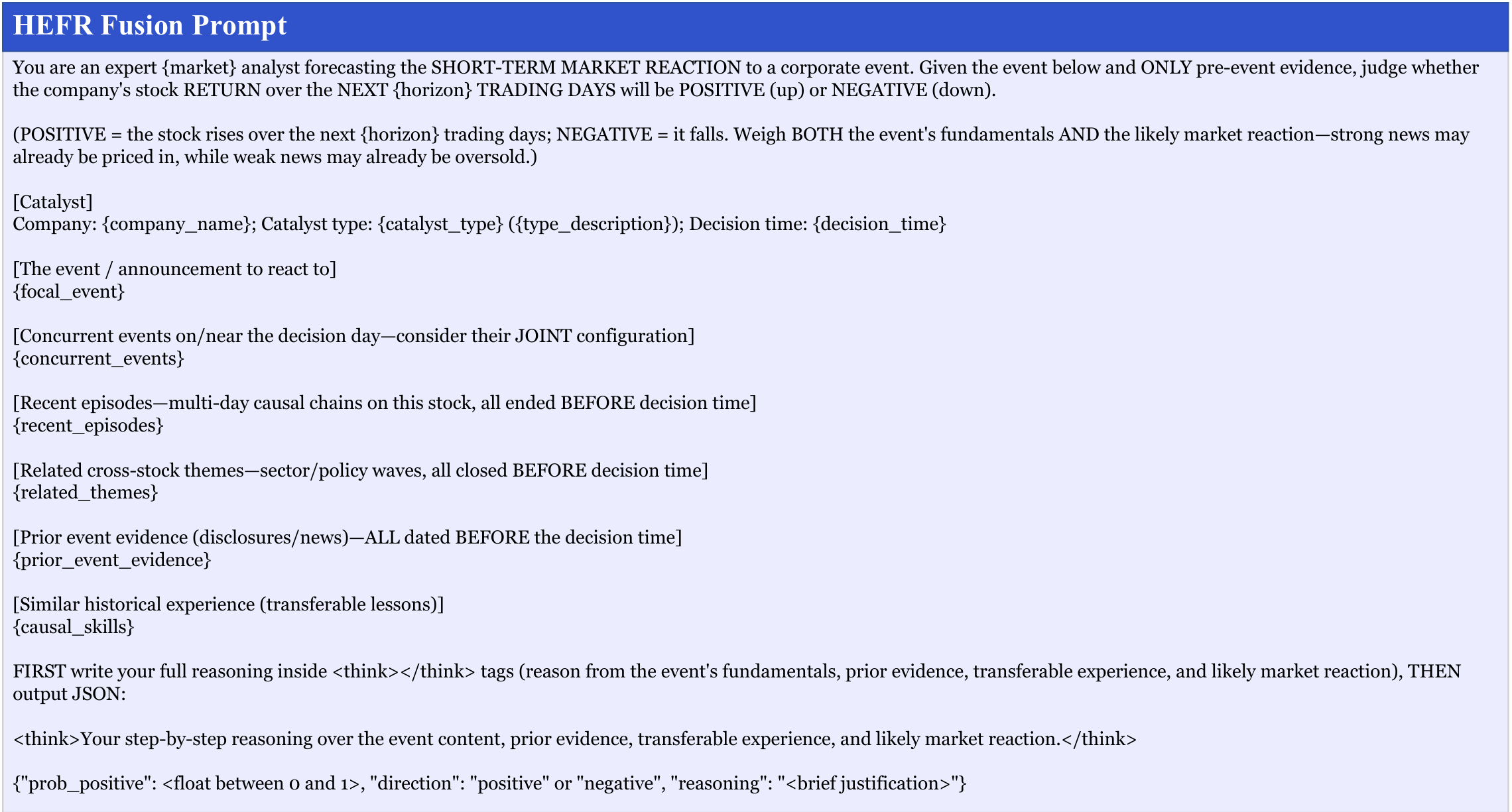}
    \caption{The heterogeneous evidence--experience fusion prompt template for HEFR.}
    \label{fig:hefr-fusion-prompt}
    \vspace{1mm}
    \includegraphics[width=1.0\textwidth]{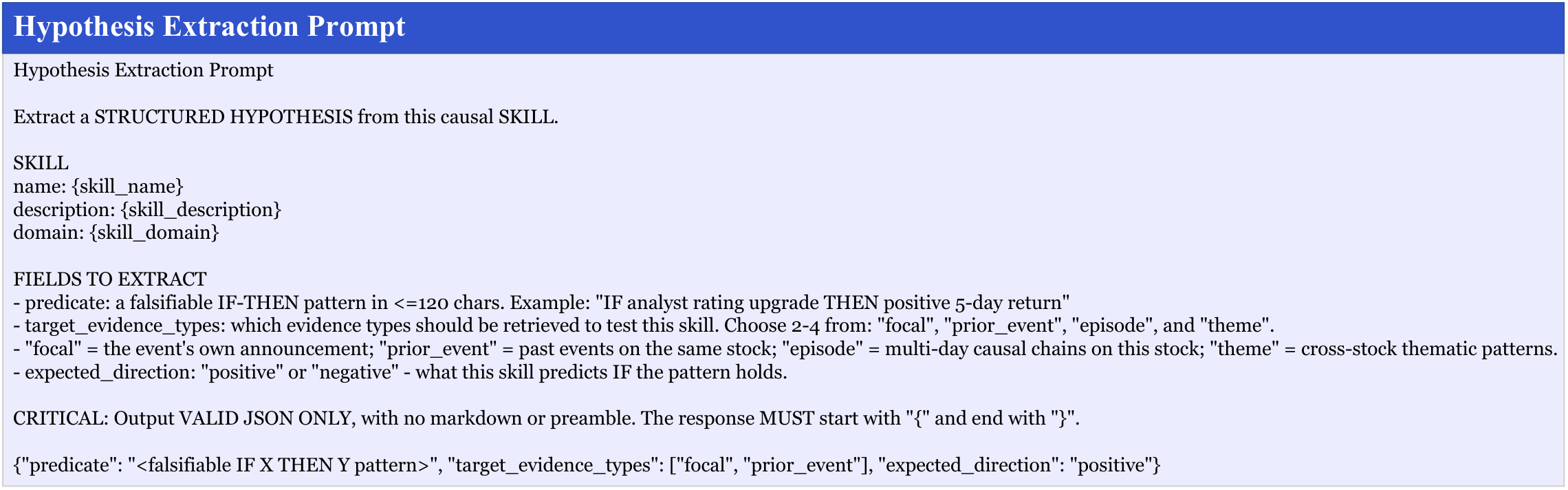}
    \caption{The SCMR prompt template for converting a case-derived skill into a structured, falsifiable hypothesis.}
    \label{fig:scmr-hypothesis-prompt}
    \vspace{1mm}
    \includegraphics[width=1.0\textwidth]{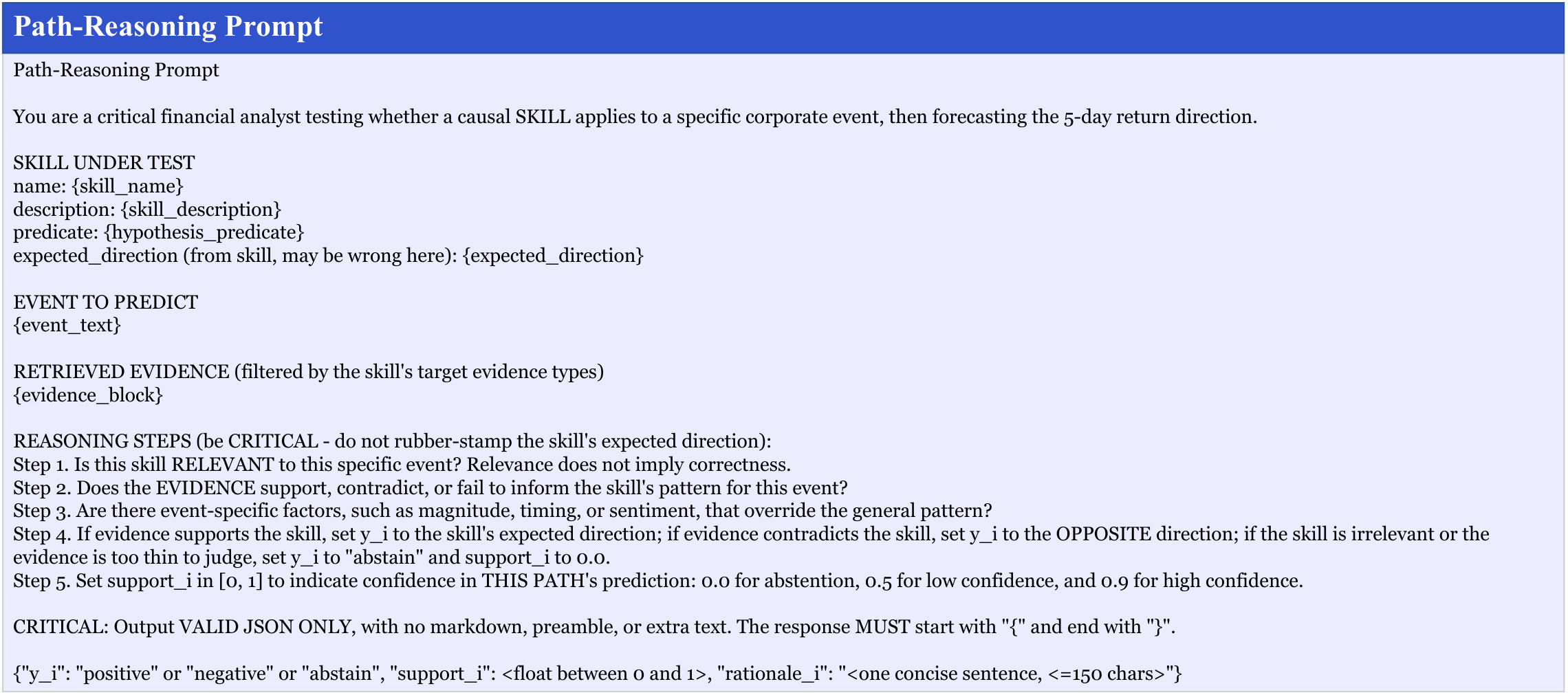}
    \caption{The SCMR prompt template for skill-conditioned evidence validation and path-level forecasting.}
    \label{fig:scmr-path-prompt}
\end{figure*}

\subsection{HEFR Fusion Prompt}
\label{sec:hefr-prompt}

As shown in Figure~\ref{fig:hefr-fusion-prompt}, HEFR performs one joint
inference per catalyst over five timestamp-gated streams: the focal event, prior
events on the same stock, recent event episodes, cross-stock themes, and
retrieved case-derived skills. It also receives a separate block of concurrent
events around the decision day. The five main streams share an adaptive raw-character
budget, and the prompt asks the model to jointly weigh event fundamentals,
historical evidence, transferable experience, and likely market reaction. It
returns $(\hat p_+,\hat y,r)$, with $\hat y$ derived from $\hat p_+$ at $0.5$;
Eq.~\eqref{eq:hefr-fusion} exposes $(\hat y,r)$. If optional SCMR produces no
valid path, inference falls back to the same HEFR call.

\subsection{SCMR Hypothesis Extraction and Path-Reasoning Prompts}
\label{sec:scmr-prompt}

As shown in Figure~\ref{fig:scmr-hypothesis-prompt}, optional SCMR first
converts each selected retrieved case-derived skill into a falsifiable IF--THEN
predicate, identifies the evidence types needed to test it, and specifies its
expected direction. The selectable evidence types are the focal event, prior
events on the same stock, recent event episodes, and cross-stock themes. This
structured output controls the subsequent retrieval scope for that skill path,
forming the hypothesis-extraction stage before independent path validation and
weighted convergence.

As shown in Figure~\ref{fig:scmr-path-prompt}, the second prompt independently
tests whether each structured skill hypothesis applies to the focal event. It
asks the model to assess skill relevance, determine whether the retrieved
evidence supports or contradicts the hypothesis, and check whether
event-specific factors override the general pattern. Each path returns a
positive, negative, or abstaining decision with a support score and a concise
rationale. Valid non-abstaining paths are then aggregated using skill stability
and path support. If no valid path remains, SCMR falls back to HEFR.
Figures~\ref{fig:scmr-hypothesis-prompt}--\ref{fig:scmr-path-prompt} show the
SCMR prompts instantiated for the five-trading-day datasets; for EDT, all
horizon-specific phrases are instantiated with three trading days.

\section{\textsc{TIEM} Algorithm Details}
\label{sec:algorithm-details}

Algorithm~\ref{alg:tiem} presents the complete \textsc{TIEM} workflow. It
separates three stages that correspond to the method components in
Section~\ref{sec:method}. First, corpus-level EEH construction converts
timestamped source texts into indexed Day, Episode, and Theme evidence.
Second, timestamp-gated forecasting retrieves eligible evidence and CSM skills
at catalyst decision time, then uses optional SCMR or a single HEFR call.
Third, training-time CSM acquisition distils and stores case-derived skills only
after outcomes resolve. In the reported main configuration, EEH evidence and
CSM skills are retrieved for HEFR, while stability diagnosis, stability
reweighting, and SCMR are disabled.

\begin{algorithm*}[t]
\footnotesize
\caption{\textsc{TIEM}: Event-Evidence Construction, Timestamp-Gated Forecasting, and Skill Acquisition}
\label{alg:tiem}
\KwIn{timestamped source texts $\mathcal D$, catalyst $c=(i,T,\kappa)$, CSM $\Sigma$, encoder $\mathrm{Enc}$, LLM $f_{\mathrm{LLM}}$, retrieval budgets, raw-character budget $B_{\mathrm{char}}$, optional flag \textsc{UseSCMR}}
\KwOut{forecast direction $\hat y(c)$ and rationale $r(c)$}

\tcp{Stage 1: Corpus-level EEH construction}
$\mathcal X\leftarrow\mathrm{Chunk}(\mathcal D)$ using the configured token size and overlap\;
\ForEach{source chunk $x\in\mathcal X$}{
  Extract atomic propositions, entities, and relations with their stock, source, and timestamp\;
  Insert the resulting Day hyperedges into $H_1$\;
}
Group $H_1$ by stock and time, then form bounded nonoverlapping candidate windows\;
Detect each Episode candidate by confidence-qualified LLM voting\;
Classify and summarize each retained candidate to construct $H_2$\;
\If{$|H_2|\ge n_{\mathrm{pool}}^{\min}$}{
  Cluster Episode-summary embeddings in rolling windows using HDBSCAN with DBSCAN fallback\;
  Apply stock, industry, and size filters, then validate, name, classify, and deduplicate Themes to construct $H_3$\;
}
Encode and index searchable records in $H_1,H_2,H_3$\;

\tcp{Stage 2: Timestamp-gated catalyst forecasting}
$F(c)\leftarrow\Phi(c)$, the latest same-stock focal source with $\mathrm{ts}\le T$\;
$U(c)\leftarrow$ same-stock concurrent facts within the configured decision-day window and $\mathrm{ts}\le T$\;
$P(c)\leftarrow H_c$, prior same-stock Day evidence with $\mathrm{ts}<T$, excluding focal matches\;
$E(c)\leftarrow\mathrm{Top}_{k_2}$ same-stock Episodes with $\tau_{\mathrm e}<T$\;
$R(c)\leftarrow\mathrm{Top}_{k_3}$ Themes containing $i$ with $\tau_{\mathrm e}<T$\;
$S_k(c)\leftarrow$ relevant CSM skills whose complete ancestry has every resolve time before $T$\;
\If{\textsc{UseSCMR} and $S_k(c)\ne\varnothing$}{
  Select $S_m(c)$ with $m=\min(m_{\max},|S_k(c)|)$\;
  \ForEach{$\sigma\in S_m(c)$ in parallel}{
    Extract hypothesis $\pi_\sigma=(\varphi_\sigma,V_\sigma,d_\sigma)$ and retrieve only evidence types $V_\sigma$\;
    Obtain $(\hat y_\sigma,w_\sigma,r_\sigma)$ and discard invalid or abstaining paths\;
  }
  \If{$\mathcal V(c)\ne\varnothing$}{
    \Return the stability-support weighted direction in Eq.~\eqref{eq:csm-converge} and its leading rationale\;
  }
}
Allocate $B_{\mathrm{char}}$ to $(F,\mathrm{Exp},E,R,P)$ in HEFR priority order and redistribute unused capacity\;
Serialize the truncated streams with budget-exempt concurrent block $U(c)$ into $\widetilde C(c)$\;
$(\hat p_+(c),\hat y(c),r(c))\leftarrow f_{\mathrm{LLM}}(\Pi(\widetilde C(c)))$, with $\hat y(c)$ derived from $\hat p_+(c)$ at threshold $0.5$\;

\tcp{Stage 3: Training-time outcome-revealed CSM acquisition}
\If{training-time acquisition is enabled and $c$ has resolved}{
  Compare $\hat y(c)$ with $y(c)$ and distil reusable case-derived skills from the event, reasoning, and verdict\;
  \ForEach{new skill $\sigma_{\mathrm{new}}$}{
    Find the most similar stored skill $\sigma^*$ in the same domain\;
    Apply \textsc{Evoke}: auto-merge advantage, tags, and capped source ids, then reset the stability state if similarity reaches $\tau$; otherwise insert $\sigma_{\mathrm{new}}$ with its resolve time\;
  }
  Optionally diagnose stability on training catalysts and retire skills labelled \textsc{miss}\;
}
\Return $\hat y(c),r(c)$\;
\end{algorithm*}

\textbf{Construction and inference flow.}
EEH is constructed once per corpus and shared by its downstream catalysts,
while every retrieval is gated by the catalyst decision time. Focal and
configured concurrent records use the non-strict gate $\mathrm{ts}\le T$.
Prior Day records and retrieved Episodes and Themes satisfy
$\mathrm{ts}<T$ and $\tau_{\mathrm e}<T$, respectively. CSM acquisition
processes training catalysts chronologically. Every acquired skill stores its
source ancestry, which is unioned across automatic and explicit merges. A skill
is unavailable unless every ancestor has a verifiable resolve time before $T$;
missing, invalid, or late ancestry therefore fails closed. During evaluation,
the CSM is frozen and read-only,
so test catalysts do not update one another and may be forecast in parallel.

\textbf{Computational complexity.}
Let $N_{\mathrm{ch}}$ be the number of source chunks, $W_2$ the number of
Episode candidate windows, $W_2^+$ the retained candidates, $W_3^+$ the
structurally valid Theme clusters, $M_D$ the number of same-domain skills,
$d$ the embedding dimension, and $m\le m_{\max}$ the selected SCMR paths.
EEH construction makes one extraction call per uncached source chunk, up to
$N_{\mathrm{ep}}W_2$ Episode-detection calls, two additional calls per retained
Episode for classification and summarization, and up to two calls per valid
Theme candidate for validation/naming and classification. Clustering cost is
implementation-dependent and is incurred once per corpus. Per catalyst, HEFR
uses fixed-count retrievals, $O(B_{\mathrm{char}})$ serialization, and one
logical LLM call. Optional uncached SCMR adds at most one hypothesis-extraction
and one path-reasoning call per selected skill, for $O(m)$ total path work that
can execute in parallel. Outcome-revealed acquisition uses one distillation
call and an exact same-domain similarity scan costing $O(M_Dd)$ before EVOKE.
In practice, runtime is dominated by LLM and embedding calls rather than by
context serialization.

\setcounter{section}{3}
\section{Dataset Details}
\label{sec:dataset-details}

We use five event-driven financial forecasting datasets spanning the Chinese
and U.S. equity markets and multiple time periods. Each dataset is normalized
into catalyst instances containing a security, catalyst type, decision time,
resolve time, event text, and forward-return direction. EDT uses a three-trading-day
forecast horizon, while the other datasets use five trading days.

\textbf{Astock.}
Astock~\cite{astock} is an event-driven Chinese A-share dataset covering July
2018 to December 2020. It contains five catalyst types: earnings, large moves,
mergers and acquisitions, ratings, and regulatory events, together with related
corporate announcements and news. Each instance is labelled by the five-trading-day
return direction following the event.

\textbf{CMIN-US.}
CMIN-US is derived from CMIN-Dataset~\cite{cmin} and covers U.S. equity events
from 2018 to 2021. It pairs corporate announcements and market news with their
subsequent stock returns and includes earnings, large moves, mergers and
acquisitions, ratings, and regulatory events. Each instance uses the
five-trading-day forward-return direction as its label.

\textbf{EDT.}
EDT is derived from TradeTheEvent~\cite{edt}, an event-driven U.S. equity
dataset covering September 2020 to April 2021. It links corporate-event texts
to market reactions across multiple event types. Following TradeTheEvent's
native price labels, available for up to three trading days, EDT uses the
three-trading-day forward-return direction as its label.

\textbf{CSMD.}
CSMD is derived from the CSMD50 and CSMD300 subsets of
CSMD-LightQuant~\cite{csmd}, covering Chinese equity events from 2021 to 2024.
It contains company-related event texts, event times, and corresponding market
returns across multiple catalyst types. Each instance is labelled by the
five-trading-day return direction following the event.

\textbf{FinPURE.}
FinPURE is our recent-period A-share earnings dataset, comprising earnings
announcements released between October 2024 and January 2025. Each instance
contains the announcement text, decision time, resolve time, and a five-trading-day
price window. Its forward return and direction label can be recomputed directly
from the stored price window, supporting independent verification of label
correctness and temporal order. FinPURE also provides maskable security-name
and date fields for the per-model Name-Date Probe.

\textbf{Data construction and test protocol.}
For each dataset, we construct a separate EEH from its timestamped text corpus.
CSM is constructed from 3,000 resolved catalysts randomly sampled from Astock
and excluded from the final test set. For evaluation, we randomly sample 128
instances from each dataset's test-candidate pool to form its final test set,
yielding 640 test instances in total. These five test sets do not overlap with
the CSM-construction instances and are used to evaluate \textsc{TIEM}, all
baselines, and both backbone models. EEH indexes event text without using the
ground-truth outcomes of test instances, and timestamp gates exclude
post-decision evidence when the record time is verifiable.

\section{Baseline Details}
\label{sec:baseline-details}

The baselines fall into three families: direct prompting, memory-augmented
methods, and retrieval-augmented methods. They share test instances, targets,
output format, and backbone-specific decoding, while evidence access follows
each method's retrieval or memory mechanism.
Direct prompting relies on the LLM itself, memory-augmented methods use persistent
historical memory, and retrieval-augmented methods obtain relevant information
from external text or structured knowledge.

\textbf{LLM Zero-shot.}
This method directly provides the catalyst query and prediction instruction to the
LLM without demonstrations, external knowledge, or additional training. The
model generates its prediction solely from its parametric knowledge.

\textbf{CoT.}
Chain-of-thought prompting~\cite{cot-prompting} elicits intermediate reasoning
steps that decompose a complex task into a sequential analysis. The model
produces its final prediction after completing the step-by-step reasoning.

\textbf{MemGPT.}
We use a Letta v0.16.8 archival-memory adapter inspired by the tiered memory
and context-management pattern of MemGPT~\cite{memgpt}. It retrieves historical
blocks from persistent memory for the prediction prompt, rather than reproducing
the complete original MemGPT system.

\textbf{Mem0.}
Mem0~\cite{mem0} extracts information with long-term value from interactions
and maintains persistent memory through addition, update, consolidation, and
deletion. Relevant memories are retrieved and inserted into the model context
for subsequent tasks.

\textbf{A-MEM.}
A-MEM~\cite{a-mem} represents memories as structured notes containing semantic
content, keywords, and tags, and automatically establishes links among them.
Adding a new memory can trigger dynamic updates to existing memories and their
relationships.

\textbf{Vanilla RAG.}
Vanilla RAG~\cite{rag-lewis} encodes external documents as dense vectors and
retrieves relevant text according to semantic similarity between the query and
documents. The retrieved text is supplied to the LLM as additional context for
prediction.

\textbf{HippoRAG.}
HippoRAG~\cite{hipporag} extracts entities and relations from text to construct
a knowledge graph, then performs associative propagation using personalized
PageRank. This mechanism connects related information distributed across
documents and supports multi-hop retrieval.

\textbf{GraphRAG.}
GraphRAG~\cite{graphrag} extracts entities and relations from a text corpus and
produces hierarchical summaries of graph communities. It retrieves local
entity-relation context or global community information to provide structured
context for the query.

\textbf{LightRAG.}
LightRAG~\cite{lightrag} constructs an entity-relation graph and combines
low-level entity retrieval with high-level thematic retrieval. The retrieved
graph structures and text are jointly supplied to the LLM as generation
context.

\textbf{HyperGraphRAG.}
HyperGraphRAG~\cite{hypergraphrag} represents higher-order relations among
multiple entities with hyperedges and organizes textual knowledge as a
hypergraph. Retrieval exploits these higher-order connections to obtain
query-relevant entities, relations, and textual context.

\section{Evaluation Details}
\label{sec:evaluation-details}

We evaluate event-driven financial forecasting along five dimensions:
predictive performance, contamination sensitivity, cross-dataset shift performance,
final-forecast accuracy--token trade-off, and cross-backbone prediction consistency
on the evaluated pair. We map task labels $-1/+1$ to $0/1$ for evaluation. For
$N$ test instances, let $y_i,\hat y_i\in\{0,1\}$ denote the ground-truth and predicted labels.

\textbf{Predictive performance.}
Accuracy measures the fraction of correct predictions:
\begin{equation}
\mathrm{Acc}=\frac{TP+TN}{TP+TN+FP+FN}.
\end{equation}
The Matthews Correlation Coefficient uses all four entries of the confusion
matrix:
\begin{equation}
\begin{aligned}
\mathrm{MCC}&=\frac{TP\,TN-FP\,FN}{D},\\
D^2&=(TP+FP)(TP+FN)\\
&\quad\cdot(TN+FP)(TN+FN).
\end{aligned}
\end{equation}
We set MCC to zero when its denominator is zero. For class-balanced
performance, we compute
\begin{equation}
\begin{aligned}
\mathrm{F1}_{+}&=\frac{2TP}{2TP+FP+FN},\\
\mathrm{F1}_{-}&=\frac{2TN}{2TN+FP+FN}.
\end{aligned}
\end{equation}
and report
\begin{equation}
\mathrm{MacroF1}=\frac{\mathrm{F1}_{+}+\mathrm{F1}_{-}}{2},
\end{equation}
with a zero value for a class-specific term whose denominator is zero. Acc
and macro F1 are reported in percent, MCC as a coefficient, and higher values
are better for all three.

\textbf{Contamination sensitivity.}
The Name-Date Probe uses the nested inputs
$E_{\mathrm{nd}}\subset E_{\mathrm{nc}}\subset E_{\mathrm{full}}$ defined in
Section~\ref{sec:task}. MemBase measures accuracy from the security identity
and date alone:
\begin{equation}
\mathrm{MemBase}=\mathrm{Acc}(f;E_{\mathrm{nd}}).
\end{equation}
We compare it with the majority-class base rate
\begin{equation}
b=\max\left\{\frac{1}{N}\sum_{i=1}^{N}y_i,
1-\frac{1}{N}\sum_{i=1}^{N}y_i\right\}.
\end{equation}
Event-content contribution is the paired accuracy difference
\begin{equation}
\Delta_{\mathrm{content}}=
\mathrm{Acc}(f;E_{\mathrm{full}})-\mathrm{Acc}(f;E_{\mathrm{nc}}).
\end{equation}
A positive value indicates predictive information in the focal-event text
beyond identity, date, recent statistics, and prior headlines. MemBase is an
audit signal rather than proof of contamination, and
$\Delta_{\mathrm{content}}$ measures content dependence rather than excluding
all possible shortcuts.

\textbf{Cross-dataset shifts.}
For each method and backbone, reference accuracy averages Astock and FinPURE,
whereas cross-dataset-shift accuracy averages CMIN-US, EDT, and CSMD, denoted
$\mathcal D_{\mathrm{shift}}$:
\begin{equation}
\mathrm{Acc}_{\mathrm{ref}}=
\frac{\mathrm{Acc}_{\mathrm{Astock}}+\mathrm{Acc}_{\mathrm{FinPURE}}}{2},
\end{equation}
\begin{equation}
\begin{aligned}
\mathrm{Acc}_{\mathrm{shift}}=\frac{1}{3}\big(&
\mathrm{Acc}_{\mathrm{CMIN\text{-}US}}+\mathrm{Acc}_{\mathrm{EDT}}\\
&+\mathrm{Acc}_{\mathrm{CSMD}}\big).
\end{aligned}
\end{equation}
Worst-case shift accuracy and the robustness ratio are
\begin{equation}
\mathrm{Acc}_{\mathrm{worst}}=
\min_{d\in\mathcal D_{\mathrm{shift}}}\mathrm{Acc}_{d},\qquad
\mathrm{RR}=\frac{\mathrm{Acc}_{\mathrm{shift}}}
{\mathrm{Acc}_{\mathrm{ref}}}.
\end{equation}
Higher values are better. Because a ratio can favor a method with weak
reference accuracy, we interpret RR jointly with absolute shift average and
worst-case shift accuracy.

\textbf{Final-forecast accuracy--token trade-off.}
Let $T_i^{\mathrm{p}}$ and $T_i^{\mathrm{c}}$ denote final-forecast prompt and completion
tokens for query $i$; upstream operations are excluded. Their average is
\begin{equation}
\overline T=\frac{1}{N}\sum_{i=1}^{N}
\left(T_i^{\mathrm{p}}+T_i^{\mathrm{c}}\right).
\end{equation}
Let $\overline{\mathrm{Acc}}_m$ be method $m$'s macro-average accuracy over
the five benchmarks. The log-adjusted trade-off score is
\begin{equation}
E_{\log}(m)=\frac{\overline{\mathrm{Acc}}_m}{\ln(\overline T_m)}.
\end{equation}
Using the corresponding Zero-shot result, indexed by $0$, marginal accuracy
gain per 1k extra tokens is
\begin{equation}
\begin{aligned}
E_{\mathrm{marg}}(m)&=1000\,
\frac{\overline{\mathrm{Acc}}_m-\overline{\mathrm{Acc}}_0}
{\overline T_m-\overline T_0},\\
&\hspace{28mm}\overline T_m>\overline T_0.
\end{aligned}
\end{equation}
Lower average final-forecast token use is better, whereas higher trade-off scores are
better. Pareto frontiers contain methods for which no displayed method
achieves both higher accuracy and lower token use.

\textbf{Cross-backbone prediction consistency on the evaluated pair.}
Let $\hat y_i^{(1)}$ and $\hat y_i^{(2)}$ be predictions from the two
backbones on the same instances. Prediction agreement is
\begin{equation}
\mathrm{Agreement}=\frac{1}{N}\sum_{i=1}^{N}
\mathbf{1}[\hat y_i^{(1)}=\hat y_i^{(2)}].
\end{equation}
If $q_1$ and $q_2$ are their respective positive-prediction rates, expected
agreement is $p_e=q_1q_2+(1-q_1)(1-q_2)$. Cohen's kappa is
\begin{equation}
\kappa=\frac{p_o-p_e}{1-p_e},
\end{equation}
where $p_o$ is observed agreement. The dual-correct rate is
\begin{equation}
\mathrm{DualCorrect}=\frac{1}{N}\sum_{i=1}^{N}
\mathbf{1}[\hat y_i^{(1)}=y_i]\,
\mathbf{1}[\hat y_i^{(2)}=y_i].
\end{equation}
All three measures are first computed per dataset and then macro-averaged over
the five benchmarks; higher values indicate greater prediction consistency on the
evaluated backbone pair.

\textbf{Statistical reporting.}
Displayed confidence intervals use 1,000 bootstrap resamples and percentile
$95\%$ intervals. Cross-dataset intervals resample independently within each
dataset before macro-averaging. The content contribution uses paired resampling
of the same instances. NDP interpretation relies on the reported confidence
intervals and observed gaps rather than bootstrap tail probabilities.

\begin{table*}[t]
\caption{Implementation hyperparameters for \textbf{TIEM}. Core EEH, CSM, and
HEFR settings are listed, while optional stability diagnosis, reweighting, and
SCMR settings are disabled in the main configuration.}
\label{tab:implementation-hparams}
\centering
\fontsize{9pt}{8pt}\selectfont
\setlength{\tabcolsep}{5pt}
\renewcommand{\arraystretch}{0.96}
\begin{tabular*}{\textwidth}{@{\extracolsep{\fill}}llc@{}}
\toprule
\textbf{Module} & \textbf{Hyperparameter} & \textbf{Value} \\
\midrule
EEH & HEFR retrieval budgets $(k_1,k_2,k_3)$ & $(6,3,2)$ \\
EEH & Day candidate multiplier $\xi_1$ & $8$ \\
EEH & Source chunk size and overlap & $(512,64)$ tokens \\
EEH & Episode span $\Delta_{\mathrm{ep}}$ & $14$ days (EDT: $30$) \\
EEH & Episode child bounds $(n_{\mathrm{ep}}^{\min},n_{\mathrm{ep}}^{\max})$ & $(2,7)$ \\
EEH & Episode source-prefix length & $80$ characters \\
EEH & Episode votes $(N_{\mathrm{ep}},v_{\mathrm{ep}})$ & $(3,1)$ \\
EEH & Episode confidence $\theta_{\mathrm{ep}}$ & $0.70$ \\
EEH & Theme window and step $(\Delta_{\mathrm{th}},\delta_{\mathrm{th}})$ & $(90,30)$ days \\
EEH & Theme minima and cluster cap $(n_{\mathrm{st}}^{\min},n_{\mathrm{ind}}^{\min},n_{\mathrm{cl}}^{\max})$ & $(2,1,20)$ \\
EEH & HDBSCAN minimum size, method, and metric & $(2,\mathrm{eom},\ell_2)$ \\
EEH & DBSCAN fallback $(\varepsilon,\mathrm{min\_samples})$ & $(0.85,2)$ \\
EEH & Theme naming-summary cap & $10$ Episodes \\
EEH & Theme deduplication threshold and total cap & $(0.85,200)$ \\
EEH & Theme pool minimum and confidence $(n_{\mathrm{pool}}^{\min},\theta_{\mathrm{th}})$ & $(50,0.60)$ \\
EEH & Concurrent same-day fact cap $n_U$ & $12$ \\
\midrule
CSM & Source-id cap $n_p$ and EVOKE threshold $\tau$ & $(20,0.85)$ \\
CSM & EMA retention $\lambda_a$ & $0.90$ \\
CSM & Advantage bounds $(a_{\min},a_{\max})$ & $(-1,1)$ \\
CSM & Skill-description cap $L_\psi$ & $500$ characters \\
CSM & Explicit-merge weight $\lambda_m$ & $0.50$ \\
CSM & Stability sample cap and minimum $(K_{\mathrm{stab}},K_{\min})$ & $(20,5)$ \\
CSM & Stability thresholds $(\eta,\theta_l,\theta_h)$ & $(0.70,0.30,0.70)$ \\
CSM & Query prefix length $L_q$ & $400$ characters \\
CSM & Domain boost, retrieval size, and similarity $(\gamma,k,\mu)$ & $(0,5,0.30)$ \\
CSM & Maximum SCMR paths $m_{\max}$ & $3$ \\
CSM & SCMR per-path retrieval budgets $(k_1^{\mathrm{path}},k_2,k_3)$ & $(3,3,2)$ \\
CSM & Stability weights $(\alpha_{\mathrm{st}},\alpha_{\mathrm{ue}},\alpha_{\mathrm{un}},\alpha_{\mathrm{mi}})$ & $(1.2,1.0,0.7,0)$ \\
\midrule
HEFR & Focal-evidence count $k_F$ & $1$ \\
HEFR & Raw-character budget $B_{\mathrm{char}}$ & $8000$ \\
HEFR & Shares $\boldsymbol{\omega}$ in order $\mathcal A$ & $(0.20,0.30,0.18,0.12,0.20)$ \\
Global & Decoding temperature $\mathcal T$ & $0$ \\
\bottomrule
\end{tabular*}
\end{table*}

\section{Implementation Details}
\label{sec:implementation-details}

We evaluate \textbf{TIEM} and all baselines with DeepSeek-V4-Flash and GPT-5.4-mini.
All methods are evaluated on the same instances under a common target definition,
while evidence access follows each method's documented mechanism. \textbf{TIEM} uses text-embedding-3-small for EEH
retrieval and applies explicit timestamp gates, including complete-ancestry
CSM filtering; baselines filter temporally only
when record-level timestamps are available.
The main benchmark comparison uses a shared evaluation set, whereas NDP is
conducted on independent Astock and FinPURE diagnostic pools.
GPT-4o-mini extracts Days and DeepSeek-v3 constructs Episodes/Themes. All methods
use deterministic temperature-zero decoding. \textbf{TIEM} and all baselines are evaluated
three times, and Table~\ref{T1} reports the mean across the three runs. The main \textbf{TIEM} configuration uses
timestamp-gated EEH retrieval, complete-ancestry CSM, and HEFR joint fusion; stability
diagnosis, stability reweighting, and SCMR are optional and disabled in the
reported experiments. Table~\ref{tab:implementation-hparams}
reports the parameters, which remain fixed across datasets and backbones unless
otherwise specified.

\section{Temporal Validation}
\label{sec:temporal-validation}

\subsection{Complete-Ancestry Temporal Availability Audit}
\label{sec:complete-ancestry-audit}

To verify that \textsc{TIEM} accesses only information available at each
forecasting decision, we audit every evidence record that enters the prediction
context across 1,280 cases from one complete audit run. Focal and concurrent evidence must
satisfy $\mathrm{ts}\le T_c$, while prior evidence, Episodes, and Themes must end
strictly before $T_c$. For every retrieved CSM skill $\sigma$, we additionally
traverse its complete provenance ancestry and require
\begin{equation}
\mathrm{resolve\_time}(u)<T_c,\qquad
\forall u\in\mathrm{Ancestors}(\sigma).
\label{eq:complete-ancestry-audit}
\end{equation}
Skills with missing or invalid ancestry metadata, or with any ancestor that
fails this constraint, are excluded from retrieval.

\begin{table*}[t]
\caption{Temporal availability audit of retrieved evidence and complete CSM
ancestry. \textsc{Late} denotes records later than the applicable decision time;
\textsc{Unver.} denotes missing or invalid temporal metadata.}
\label{tab:complete-ancestry-audit}
\centering
\fontsize{9pt}{8pt}\selectfont
\setlength{\tabcolsep}{5pt}
\begin{tabular*}{\textwidth}{@{\extracolsep{\fill}}lrrrr@{}}
\toprule
\textbf{Audited object} & \textbf{Checked records} & \textbf{Late} & \textbf{Unverifiable} & \textbf{Affected queries} \\
\midrule
Focal evidence & 4,586 & 0 & 0 & 0 \\
Prior evidence & 3,332 & 0 & 0 & 0 \\
Concurrent evidence & 2,842 & 0 & 0 & 0 \\
Episodes & 920 & 0 & 0 & 0 \\
Themes & 290 & 0 & 0 & 0 \\
Retrieved CSM skills & 6,040 & 0 & 0 & 0 \\
CSM provenance ancestors & 16,301 & 0 & 0 & 0 \\
\bottomrule
\end{tabular*}
\end{table*}

Table~\ref{tab:complete-ancestry-audit} shows that all focal, prior,
concurrent, Episode, and Theme records satisfy their corresponding temporal
constraints. The 6,040 retrieved CSM skills contain 16,301 verifiable provenance
ancestors, all of which resolve strictly before their corresponding decision
times. No late or unverifiable ancestry and no affected query are observed,
confirming that complete-ancestry filtering preserves the temporal availability
of the historical skills used by \textsc{TIEM}.

\subsection{Temporal-Gate Ablation}
\label{sec:temporal-gate-ablation}

\begin{table}[t]
\caption{Temporal-gate ablation on Astock. \textit{w/o} Time Gate removes the
decision-time constraints from multi-tier evidence retrieval and CSM skill
invocation, whereas TIEM (Full) retains the complete temporal-gating mechanism.
Acc and F1 are in \%; MCC is a coefficient; and \textbf{best} is in bold.}
\label{tab:temporal-gate-ablation}
\centering
\fontsize{9pt}{8pt}\selectfont
\setlength{\tabcolsep}{1.8pt}
\begin{tabular}{l|ccc|ccc}
\toprule
\multirow{2}{*}{\textbf{Method}} & \multicolumn{3}{c|}{\textbf{DeepSeek-V4-Flash}} & \multicolumn{3}{c}{\textbf{GPT-5.4-mini}} \\
\cmidrule(lr){2-4} \cmidrule(lr){5-7}
& \textbf{Acc$\uparrow$} & \textbf{MCC$\uparrow$} & \textbf{F1$\uparrow$} & \textbf{Acc$\uparrow$} & \textbf{MCC$\uparrow$} & \textbf{F1$\uparrow$} \\
\midrule
\textit{w/o} Time Gate & 59.11 & 0.18 & 58.84 & 55.73 & 0.11 & 55.68 \\
\rowcolor{grpOurs} \textbf{TIEM (Full)} & \textbf{65.62} & \textbf{0.31} & \textbf{65.59} & \textbf{65.62} & \textbf{0.31} & \textbf{65.32} \\
\bottomrule
\end{tabular}
\end{table}

Table~\ref{tab:temporal-gate-ablation} compares Full TIEM with the variant that
removes temporal gating. TIEM (Full) achieves higher Acc, MCC, and F1 on both
backbones, showing that temporal gating helps preserve consistency among the
current event, historical events, and resolved outcomes. Without temporal
gating, multi-tier retrieval can mix event evidence from different forecasting
stages, weakening both the evolutionary relations organized by EEH and the
correspondence between CSM skills and the current catalyst. Full TIEM constrains
focal, prior, concurrent, Episode, Theme, and CSM evidence by the decision time,
providing HEFR with temporally ordered and mutually coordinated evidence and
experience, thereby yielding consistent gains in directional accuracy,
predictive correlation, and class-balanced performance.

\section{Case Study}
\label{sec:case-study}

\begin{figure*}[t]
\centering
\includegraphics[width=1\linewidth]{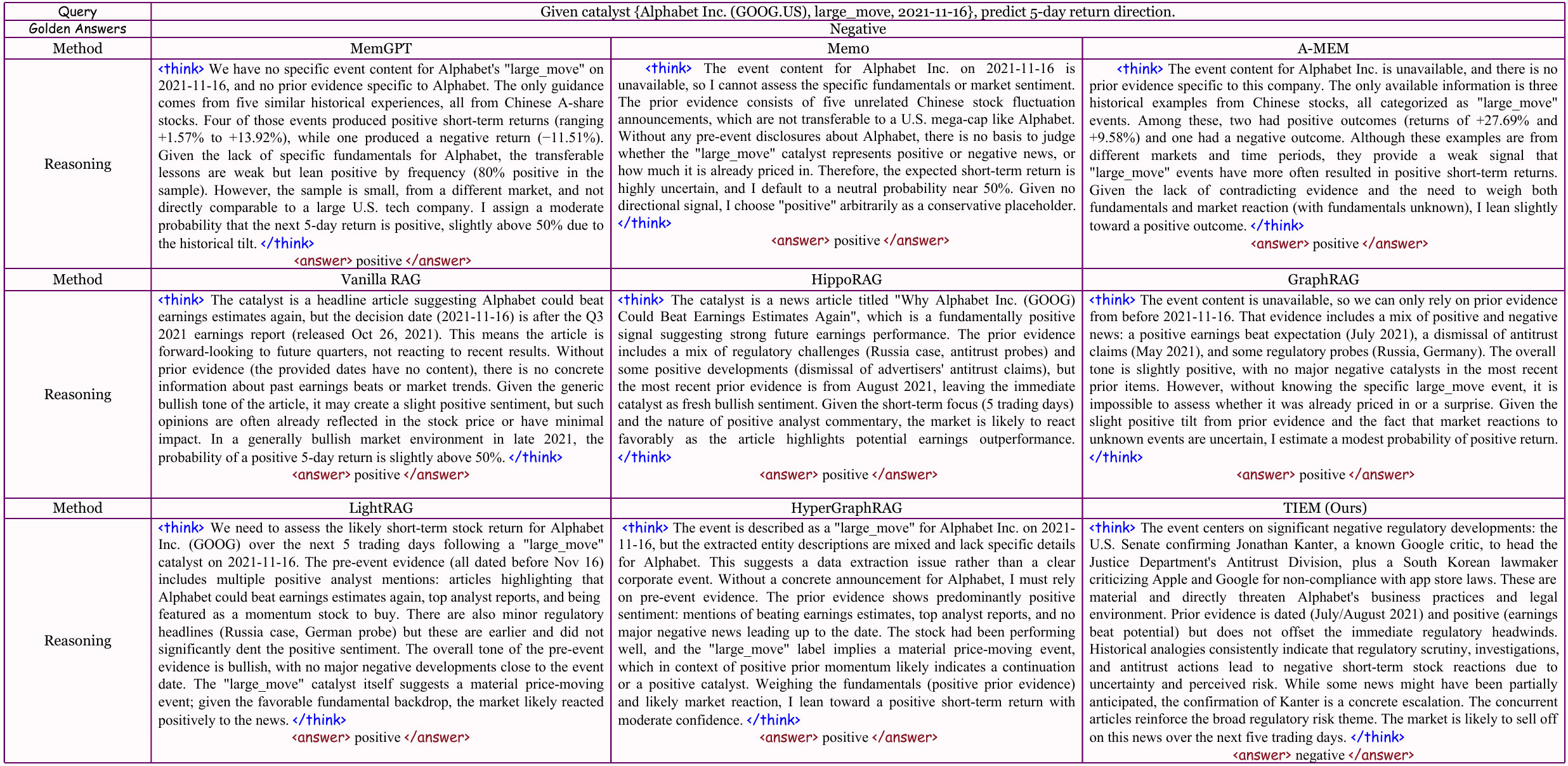}
\caption{Alphabet (GOOG.US) antitrust catalyst case study under DeepSeek-V4-Flash, comparing \textbf{TIEM} with memory and retrieval baselines for five-day return-direction prediction.}
\label{fig:case_study}
\end{figure*}

\textbf{Case-level evidence.}
Fig.~\ref{fig:case_study} presents one illustrative Alphabet (GOOG.US) case
from CMIN-US. The catalyst is an antitrust-related large-move event dated
November 16, 2021, and the five-trading-day ground-truth direction is negative.
All three memory-augmented and five retrieval-augmented baselines shown in the
figure predict positive; only \textbf{\textsc{TIEM}} predicts the correct
negative direction.

The memory baselines do not receive the focal event body and instead rely on
cross-market historical memories. MemGPT and A-MEM form a weak positive tilt
from the retrieved outcomes, while Mem0 selects a positive prediction under
high uncertainty. The retrieval baselines obtain external context but fail to
locate the decision-day regulatory event. Vanilla RAG and HippoRAG emphasize
an older earnings-expectation headline together with unrelated-company news;
GraphRAG reasons from earlier mixed evidence; and LightRAG and HyperGraphRAG
are influenced by broad or mixed graph entities and earlier positive analyst
context. In each case, the older positive context dominates the final
prediction.

\textbf{\textsc{TIEM}} instead retrieves the decision-day event document
through the EEH focal Day stream and supplies its same-day regulatory facts
through the separate concurrent stream. These facts include the U.S. Senate's
confirmation of Google critic Jonathan Kanter to lead the Justice Department's
Antitrust Division and a South Korean lawmaker's criticism of Apple and
Google's app-store compliance. The prior Day stream separately retains the
July and August earnings-expectation and analyst context. CSM retrieves case-derived
skills dominated by negative patterns of regulatory scrutiny and legal risk.
With SCMR disabled, a single HEFR call jointly weighs the focal event,
same-day facts, prior evidence, and case-derived skills, determining that the
immediate regulatory escalation outweighs the older positive context and
predicting a negative five-day return direction. This example illustrates the
system's reasoning path, but is not aggregate performance evidence. It shows
how the focal event and relevant historical evidence are jointly organized
for this case.

\section{Limitations}
\label{sec:limitations}

Although \textsc{TIEM} unifies event evidence and historical skills, it has three
boundaries. First, EEH relies on timestamped text with source metadata and LLM
extraction of facts and higher-level records. Date-only records use dataset-level
end-of-day semantics, while sparse or incomplete timestamps can weaken
retrieval. Second, CSM acquires skills only after outcomes resolve, preserves
complete ancestry through merges, and fails closed when source timing is invalid;
novel events, regime changes, and rare types may still receive limited support.
Third, HEFR fuses
focal, concurrent, prior, and experience streams under finite context, fixed
retrieval counts, and initial budgets with surplus redistribution. Method-specific evidence access prevents
isolating one component effect or establishing trading utility, universal shift
invariance, or complete contamination exclusion.

\section{Future Work}
\label{sec:future-work}

Future work mainly includes the following directions. First, EEH construction
can be strengthened through availability-time validation, source-consistency
checks, extraction-confidence estimation, and multi-sample consistency. An
incremental construction and local-revision mechanism could also incorporate
new event texts without rebuilding the complete hypergraph. Second, CSM can be
made more adaptive to novel events and changing market regimes through
cross-domain skill initialization, time-aware retrieval, periodic
revalidation, and active retirement, while improving skill transfer for rare
catalyst types. Third, retrieval counts, context budgets, and reasoning depth
can be adjusted according to query difficulty, evidence sufficiency, and
agreement across evidence streams. This direction also enables systematic
evaluation of the optional stability diagnosis and SCMR, using multi-path
validation for complex events while retaining compact joint HEFR reasoning
when the available evidence is sufficient.

\section{Ethical Considerations}
\label{sec:ethical-considerations}

We evaluate \textsc{TIEM} offline. It is neither investment advice nor an
automated trading system, and its predictions
should not be used for financial decisions without human oversight and
appropriate risk controls. FinPURE is built from corporate earnings announcements
and corresponding market-price windows and does not involve human-subject
research. FinPURE and the accompanying code will be released in compliance with
source licenses and applicable redistribution requirements for research artifacts
and derived resources used in this study.

\end{document}